\documentclass[aps,showpacs,twocolumn,amsmath,amssymb,amsfonts,eqsecnum]{revtex4-1}
\usepackage{graphicx}
\usepackage{color}
\usepackage{natbib}
\usepackage{epstopdf}
\usepackage{hyperref}

\newcommand\eq[1]{Eq.~(\ref{#1})}

\begin{document}
\title{
Manipulation and amplification of the Casimir force\\ through surface fields using helicity}
\author{Daniel Dantchev$^{1,2}$\thanks{e-mail:
daniel@imbm.bas.bg}, and  Joseph Rudnick$^{1}$\thanks{e-mail:
jrudnick@physics.ucla.edu}} \affiliation{ $^1$ Department of Physics and Astronomy, UCLA, Los Angeles,
California 90095-1547, USA,\\$^2$Institute of
Mechanics - BAS, Academic Georgy Bonchev St. building 4,
1113 Sofia, Bulgaria 
}
\date{\today}

\begin{abstract}
We present both exact and numerical results for the behavior of the Casimir force in $O(n)$ systems with a finite extension in one direction when the system is subjected to surface fields that induce helicity in the order parameter. We show that for such systems the Casimir force in certain temperature ranges is of the order of $L^{-2}$, both above and below the critical temperature, $T_c$, of the bulk system. An example of such a system would be one with chemically modulated bounding surfaces, in which the modulation couples directly to the system's order parameter. We demonstrate  that, depending on the parameters of the system, the Casimir force can be either {\it attractive} or {\it repulsive}. The exact calculations presented are for the one dimensional $XY$ and Heisenberg models under twisted boundary conditions resulting from  finite surface fields that differ in direction by a specified angle and the three dimensional Gaussian model with surface fields in the form of plane waves that are shifted in phase with respect to each other. Additionally, we present exact and numerical results for the mean field version of the three dimensional $O(2)$ model with finite surface fields on the bounding surfaces. We find that all significant results are consistent with the expectations of finite size scaling.
\end{abstract}
\maketitle

\section{Introduction}
\label{Section_Introduction}

Casimir forces result from, and provide insight into, the behavior of a medium confined to a restricted space, canonically the region between two plane, parallel surfaces. In the case of the electromagnetic Casimir force, the medium is the vacuum, and the underlying mechanism is the set of quantum zero point or temperature fluctuations of the electromagnetic field. The now widely-investigated critical Casimir force (CCF) results from the fluctuations of an order parameter and more generally the thermodynamics of the medium supporting that order parameter in the vicinity of a critical point. In fact, the free energy of a confined medium can mediate a Casimir force at any temperature provided its excitations are long-range correlated ones. This fact, along with the wide range of options for a mediating substance opens up a range of possibilities for the study and exploitation of the Casimir force arising from a confined medium. 

One of the principal influences on the Casimir force is the nature of the bounding surface. With respect to the CCF, published investigations have been focused, almost exclusively, on systems belonging to the Ising universality class.  On a basic level, based on the behavior of coupling in the vicinity of the surface, there are three universality classes---extraordinary (or normal), ordinary and surface-bulk (or special), ones  \cite{D86,K94,BDT2000}. Experimental investigations into the influence of surface universality classes on the Casimir force have been reported in \cite{SZHHB2007,RBM2007,HHGDB2008,GMHNHBD2009,NHC2009,NDHCNVB2011,ZAB2011}.  Most of them focus on the behavior of colloids in a critical solvent. They probe the dependence of the force between boundaries on temperature, the concentration of the components of the solvent and the relative preference of the surfaces of the colloids for the components of the solvent. For example, in \cite{GMHNHBD2009} the critical thermal noise in a solvent medium consisting of a binary liquid mixture of water and 2,6-lutidine near its lower consolute point is shown to lead to attractive or repulsive forces, depending on the relative adsorption preferences of the colloid and substrate surfaces with respect to the two components of the binary liquid mixture.  On the theoretical side, the influence of the surface fields has been studied on the case of two dimensional Ising model via exact calculation \cite{NN2008,NN2009,AM2010,NN2016},  using the variational formulation due to Mikheev and Fisher \cite{B2015,Z2012}, with the help of density-matrix renormalization-group numerical method \cite{MCD99,DMC2000,DME2000,ZMD2013}, via conformal invariance \cite{VED2013,JRT2015}, Monte Carlo methods \cite{VED2013}, and numerically using bond propagation algorithms \cite{WI2015}. The three dimensional Ising model has been studied with Monte Carlo methods in \cite{VGMD2009,H2011,VMD2011,TTD2013,VD2013,V2014,MVDD2015}, mean-field type calculations \cite{PD2004,K97,MMD2010,VDK2012,THD2015,VD2015} and renormalized local functional theory \cite{OO2012}. In general, it has been shown that the Casimir force depends on the strength of the surface fields $h_1$ and $h_2$ and that it can change sign as the magnitudes of the surface field, the thickness of the films, and the temperature of the system are varied. 

For the general case of $O(n)$ systems there is no similarly thorough classification \cite{P90}. References \cite{APP91,HK92,ZPZ98,KG99,BLF2000,GAF2001,HSD2004,KNSP2013,HNSP2014} report on studies of the Casimir force in  liquid crystals, and \cite{I86,GC99,GC2002,ZRK2004,GSGC2006,MGD2007,UBMCR2003} describe investigations for $^4$He and $^3$He--$^4$He mixtures. In the case of Helium films, however, it is generally accepted that the boundary conditions are determined, in the region where the liquid behaves as a quantum liquid, by its quantum nature and, thus, cannot be easily influenced by modification of bounding surfaces, in that there are no surface fields that couple to the order parameter in such systems. In that respect liquid crystals seem much more readily adjustable, and in particular more amenable to the influence of boundary conditions.  For example, in Ref. \cite{APP91} it is shown that director fluctuations in nematics induce long-range interactions between walls, which are attractive with symmetric boundary conditions, but may become repulsive with mixed ones. In smectics such forces are longer ranged than van der Waals ones.

 In \cite{ZPZ98}  the authors concluded that in the case of finite surface coupling, the fluctuation-induced forces for nematics are weaker than in the strong anchoring limit. In the example of three-dimensional lattice XY model with nearest neighbor interaction, it has been shown \cite{BDR2011} that  the Casimir force depends in a continuous way on the parameter $\alpha$  characterizing the so-called twisted boundary conditions when the angle between the vector order parameter at the two boundaries is $\alpha$  where $0<\alpha\le \pi$. The effect is essential; depending on $\alpha$ the force can be attractive or repulsive. By varying $\alpha$ and/or the temperature $T$ one can control both the sign and the magnitude of the Casimir force in a reversible way. Furthermore, when $\alpha = \pi $, an additional phase transition, which occurs only in finite systems, has been discovered,  associated with the spontaneous symmetry breaking of the direction of rotation of the vector order parameter through the body of the system.   

In the current article we show that the strength and the mutual orientation of surface fields---as well as structuring on the surface via chemical  or other alternations that can be described in terms of surface fields---lead to interesting and substantial modification in the behavior of the force between the confining surface. Such modification includes the change of the sign of the force, as well as non-monotonic behavior, appearance of multiple minima, of a longitudinal Casimir force,  and also an amplification of the force in regions with strong helicity effects. We will demonstrate the above with the example of few models: the one dimensional XY and Heisenberg models, the three dimensional Gaussian model and the three dimension $O(2)$ XY model. 

We start with the one-dimensional XY and Heisenberg models. 

\section{1d continuum symmetry models with boundary fields} \label{sec:continuum}
\label{1d_systems}
Here we consider two one-dimensional models with continuous $O(n)$ spin symmetry: XY ($n=2$) and Heisenberg ($n=3$) chains of $N$ spins with ferromagnetic interaction $J$ between nearest-neighbor spins, 
the boundary fields ${\mathbf H}_1$ and ${\mathbf H}_2$ of which are at an angle $0\le\psi\le \pi$ with respect to each other. Obviously, such systems do not exhibit spontaneous ordering at non-zero temperatures given  their low dimension and the short range nature of the interactions between spins, as has been shown to follow rigorously from the Mermin-Wagner theorem \cite{MW66}. Nevertheless, they posses an essential singular point at $T=0$ and will, in that limit, support spontaneous order. We will demonstrate that when the boundary fields are non-zero the Casimir force, $F_{\rm Cas}$, of these systems displays very rich and interesting behavior. We also show that near $T=0$ the force has a scaling behavior and that, depending on the angle between the boundary fields and the value of the temperature scaling variable  $x\sim N k_B T/J$, this force can be {\it attractive} or {\it repulsive}. More precisely, we will establish that: 
\begin{enumerate} 
\item[i)]For low temperatures, when $x={\cal O}(1)$ and 
\begin{equation}
 \label{eq:constraint}
 N\gg J\left(\frac{1}{H_1}+\frac{1}{H_2}\right)
 \end{equation}
the leading behavior of the Casimir force can be written in the form
\begin{equation}
\label{eq:1dCas_gen}
\beta F_{\rm Cas}(T,N,{\bf H}_1,{\bf H}_2)=N^{-1}X(\psi,x),
\end{equation}
with $x$ a scaling variable  and $X$ a universal scaling function. Equation (\ref{eq:1dCas_gen}) implies that, under constraint \eq{eq:constraint}, $X_{\rm Cas}$ depends only on the scaling variable $x$ defined in (\ref{eq:scaling_variables})  and the angle $\psi$. The latter parameter effectively describes the boundary conditions on the system. Note that, unlike the Ising model, the boundary conditions depend here {\it continuously} on one parameter---in our notation $\psi$.

\item[ii)] When $x\to 0+$ the scaling function of the Casimir force becomes positive, i.e., the force turns {\it repulsive} provided that $\psi \ne 0$. In that case $X_{\rm Cas}\sim x^{-1}$ and, thus, the overall $N$-dependence of the force is of the order of $N^{-2}$.

\item[iii)] When $x\gtrsim 1$ the scaling function has a sign that depends on the sign of $\cos(\psi)$: for $0<|\psi|<\pi/2$ the force will be {\it attractive}, while for $\pi/2<|\psi|<\pi$ it will be {\it repulsive}. For $x\gg 1$ the force decays exponentially to zero. 

\item[iv)] For any $\psi$ such that $0<|\psi|<\pi/2$ the Casimir force {\it changes from attractive to repulsive} when the temperature decreases from a moderate value to zero for fixed system size, $N$. 

\item[v)] When $\psi=0$ the force is attractive for {\it any} value of the scaling variable $x$. 

\end{enumerate} 

These 1d models have been studied analytically in the case of free (frequently termed ``open'' or Dirichlet) and periodic boundary conditions \cite{F64,J67,J67b,S68,PB2011}, but we are not aware of any investigation of them in the presence of boundary fields, which are responsible for the  effects of interest in this article. 
\subsection{The 1d XY model}
\label{sec:1dXY_model}
We consider a system with the Hamiltonian
\begin{equation}
\label{eq:def_1d_Ham}
{\cal H} = -J \sum _{i=1}^{N-1} {\mathbf S}_i.{\mathbf S}_{i+1}-{\mathbf H}_1.{\mathbf S}_1-{\mathbf H}_N.{\mathbf S}_N 
\end{equation}
where ${\mathbf S}_i$, with ${\mathbf S}_i^2=1$ and ${\mathbf S}_i \in \mathbb{Z}^2$, $i=1,\cdots,N$,  are $N$ spins arranged along a straight line. The Hamiltonian can be written in  the form 
\begin{eqnarray}
\label{eq:system_angles}
{\cal H} &=& -J \sum _{i=1}^{N-1} \cos \left(\varphi _{i+1}-\varphi _i\right)
\\&&
-H_1 \cos \left(\psi _1-\varphi _1\right)-H_N
   \cos \left(\psi _N-\varphi _N\right), \nonumber
\end{eqnarray}
where the angles $\psi_1, \psi_2$ and $\varphi _1,\cdots,\varphi _N$ are measured with respect to the line of the chain which is taken to be, say, the x axis. The free energy $-\beta  F_N$ of this system is given by
\begin{equation}
\label{eq:free_energy}
\exp \left(-\beta  F_N\right)=\int_{0}^{2\pi}\exp \left(-\beta  {\cal H}\right)\ \ \prod_{i=1}^{N} \frac{d\varphi_i}{2\pi}.
\end{equation}

Performing the requisite calculations (see Appendix \ref{app:XY}) one obtains
\begin{eqnarray}
\label{eq:free_energy_calculated}
\lefteqn{\exp \left(-\beta  F_N\right)}\\&=&\sum _{k=-\infty }^{\infty } \exp \left(i k \psi\right) I_k\left(h_1\right) I_k(K){}^{N-1} I_k\left(h_N\right)\nonumber
\end{eqnarray}
where 
\begin{equation}
\label{eq:def_parameters}
\psi \equiv (\psi_1-\psi_N), K\equiv \beta J, h_1\equiv \beta H_1, h_N\equiv\beta H_N. 
\end{equation}
Note that the free energy depends only on the difference in angles, $(\psi_1-\psi_N)$,  and not on $\psi_1$ and $\psi_N$ separately. For the Casimir force in the system, i.e., for the finite size part of the total force, see \eq{tot}, one then has the {\it exact} expression
\begin{widetext}
\begin{equation}
\label{FCas}
\beta  F_{\text{Cas}}=
\frac{ 2\sum _{k=1}^{\infty } 
\cos \left[k (\psi_1-\psi_2)\right] \log \left[\frac{I_k(K)}{I_0(K)}\right]\frac{I_k\left(h_1\right)}{I_0\left(h_1\right)} \left(\frac{I_k(K)}{I_0(K)}\right)^{N-1} \frac{I_k\left(h_N\right)}{I_0\left(h_N\right)}
}{1+2 \sum _{k=1}^{\infty } \cos \left[k (\psi_1-\psi_2)\right] \frac{I_k\left(h_1\right)}{I_0\left(h_1\right)} \left(\frac{I_k(K)}{I_0(K)}\right)^{N-1} \frac{I_k\left(h_N\right)}{I_0\left(h_N\right)}}.
\end{equation}
\end{widetext}

 From here on we will be interested in the behavior of the system in the limit $\beta\gg 1$, i.e., when $T\to 0$. Obviously, when  $\beta\gg 1$ from \eq{eq:def_parameters} one has $h_1\gg 1$, $h_N\gg 1$ and $K\gg 1$, which means that in \eq{eq:free_energy_calculated} one uses the large argument asymptote of $I_{k}(z)$ for $z\gg 1$.  We will use the asymptote in the form reported in \cite{SP85}
 \begin{equation}
 \label{eq:as_form_SP}
 I_\nu(z)= \frac{e^{z-\nu^2/2z}}{\sqrt{2\pi z}}\left[1+\frac{1}{8 z}+{\cal O}\left(\frac{\nu^2}{z^2}\right)\right].
 \end{equation}
 Retaining only the first term in the above expansion, one obtains
 \begin{equation}
 \label{eq:F_Cas_scaling}
 \beta F_{\rm Cas}(x)=\frac{1}{N_{\rm eff}}X_{\rm Cas}(\psi,x,h_{\rm eff})
 \end{equation}
 where 
 \begin{equation}
  \label{eq:F_Cas_scaling}
  X_{\rm Cas}=-x\frac{ \sum _{k=1}^{\infty } k^2
     \cos \left(k \psi\right) \exp
     \left[-\frac{1}{2}
     k^2\left(h_{\rm eff}^{-1}+x\right)\right]}{
     1+2 \sum _{k=1}^{\infty } \cos \left(k \psi\right) \exp \left[-\frac{1}{2}
     k^2\left(h_{\rm eff}^{-1}+x\right)\right]},
  \end{equation}
  and
  \begin{equation}
  \label{eq:scaling_variables}
 x\equiv \frac{N_{\rm eff}}{K}, \qquad h_{\rm eff}^{-1}=h_1^{-1}+h_2^{-1}, \qquad N_{\rm eff}=N-1.
  \end{equation}
 Here, $x$ is the scaled version of the reduced temperature variable, which in systems with a non-zero transition temperature takes the form $x=t^{\nu}L$, with $t$ the reduced temperature $\propto T-T_c$, $L$ the characteristic size of the finite system and $\nu$ the correlation length exponent.  Recall that with an effective transition temperature of $T=0$ and $K \propto 1/T$, the definition in (\ref{eq:scaling_variables}) is consistent with this definition under the assumption that $\nu=1$. 
 
Obviously,  when \eq{eq:constraint} is fulfilled 
one has $x\gg h_{\rm eff}^{-1}$, and one can safely ignore $h_{\rm eff}$ in \eq{eq:F_Cas_scaling}. Then the behavior of the force is exactly as stated in \eq{eq:1dCas_gen}.
 
 The representation of $X_{\rm Cas}$ given by \eq{eq:F_Cas_scaling} is convenient for all values of $x$ except in the limit $x \ll 1$. For that limit, using the Poisson identity \eq{eq:Poisson}, one obtains 
 \begin{eqnarray}
   \label{eq:F_Cas_scaling_small_x}
  \lefteqn{X_{\rm Cas}(\psi,x,h_{\rm eff})=-\frac{x}{2
     \left(x+h_{\rm eff}^{-1}\right)}}\\
     &&+\frac{x}{2\left(x+h_{\rm eff}^{-1}\right)^2}\frac{ \sum _{n=-\infty}^{\infty} \left(2 n \pi
     +\psi\right)^2
    \exp{\left[-\frac{\left(2 n \pi +\psi\right)^2}{2
         \left(x+h_{\rm eff}^{-1}\right)}\right]}}{
      \sum _{n=-\infty}^{\infty}
     \exp{\left[-\frac{\left(2 n \pi +\psi\right)^2}{2
     \left(x+h_{\rm eff}^{-1}\right)}\right]}}. \nonumber
   \end{eqnarray}
Under the assumption that the constraint \eqref{eq:constraint} is fulfilled and given the asymptotic behavior of $X_{\rm Cas}$ from Eqs. \eqref{eq:F_Cas_scaling} and \eqref{eq:F_Cas_scaling_small_x}, we derive
\begin{equation}
\label{X_cas_ass}
X_{\rm Cas}(\psi,x)=\left\{ \begin{array}{ll}
-\frac{1}{2}+\frac{1}{2x} \psi^2+\cdots, & x\to 0+ \\
-x \cos(\psi) \exp(-x/2), & x \gg 1.
\end{array}
\right.
\end{equation}
\begin{figure}[h]
\includegraphics[width=\columnwidth]{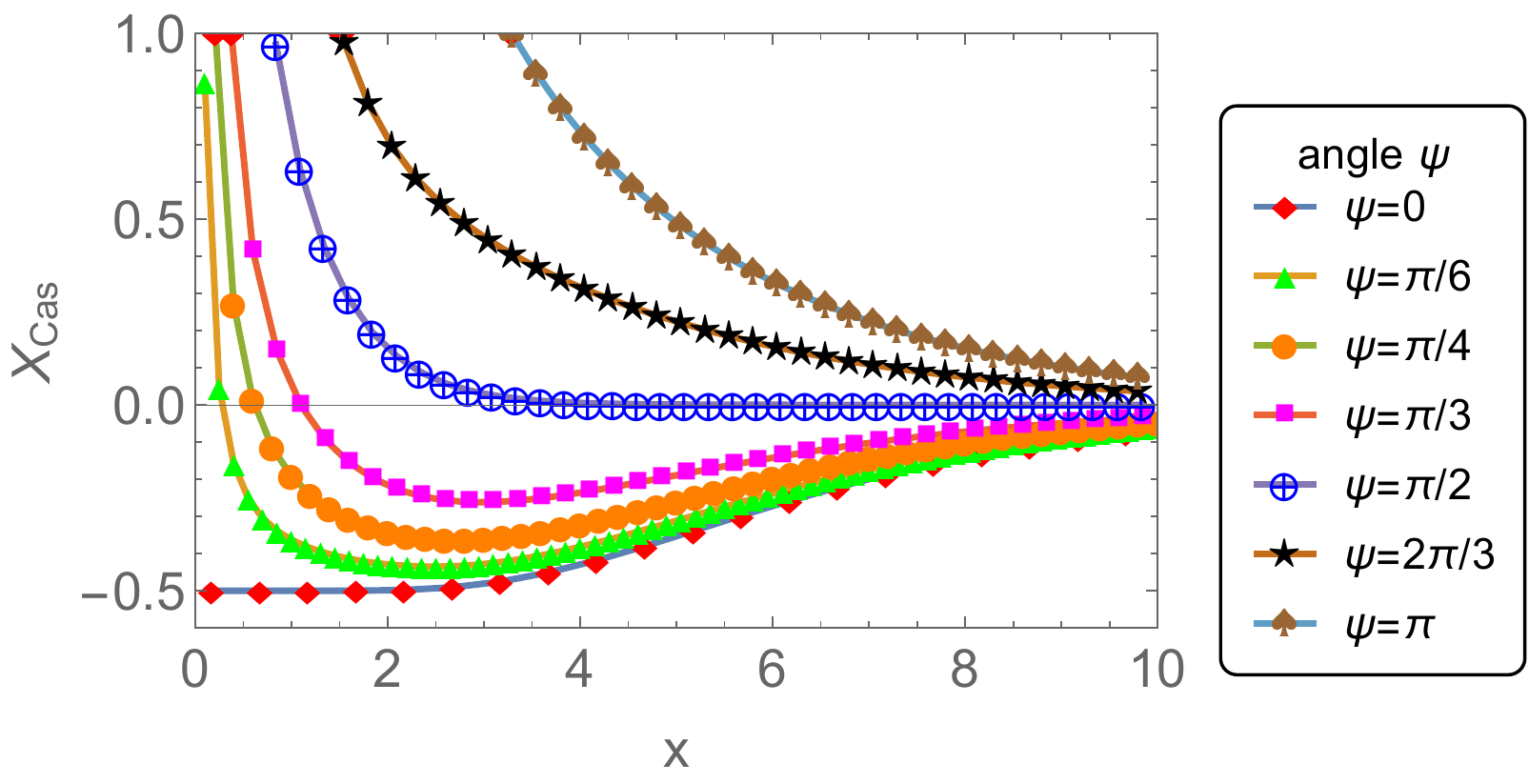}
\caption{(Color online) The scaling function $X_{\rm Cas}$ of the XY model as a function of the scaling variable $x$, see \eq{eq:scaling_variables}, for different values of the phase change $\psi$.}
\label{Fig:1d_X_Cas}
\end{figure}
\begin{figure}[h]
\includegraphics[width=\columnwidth]{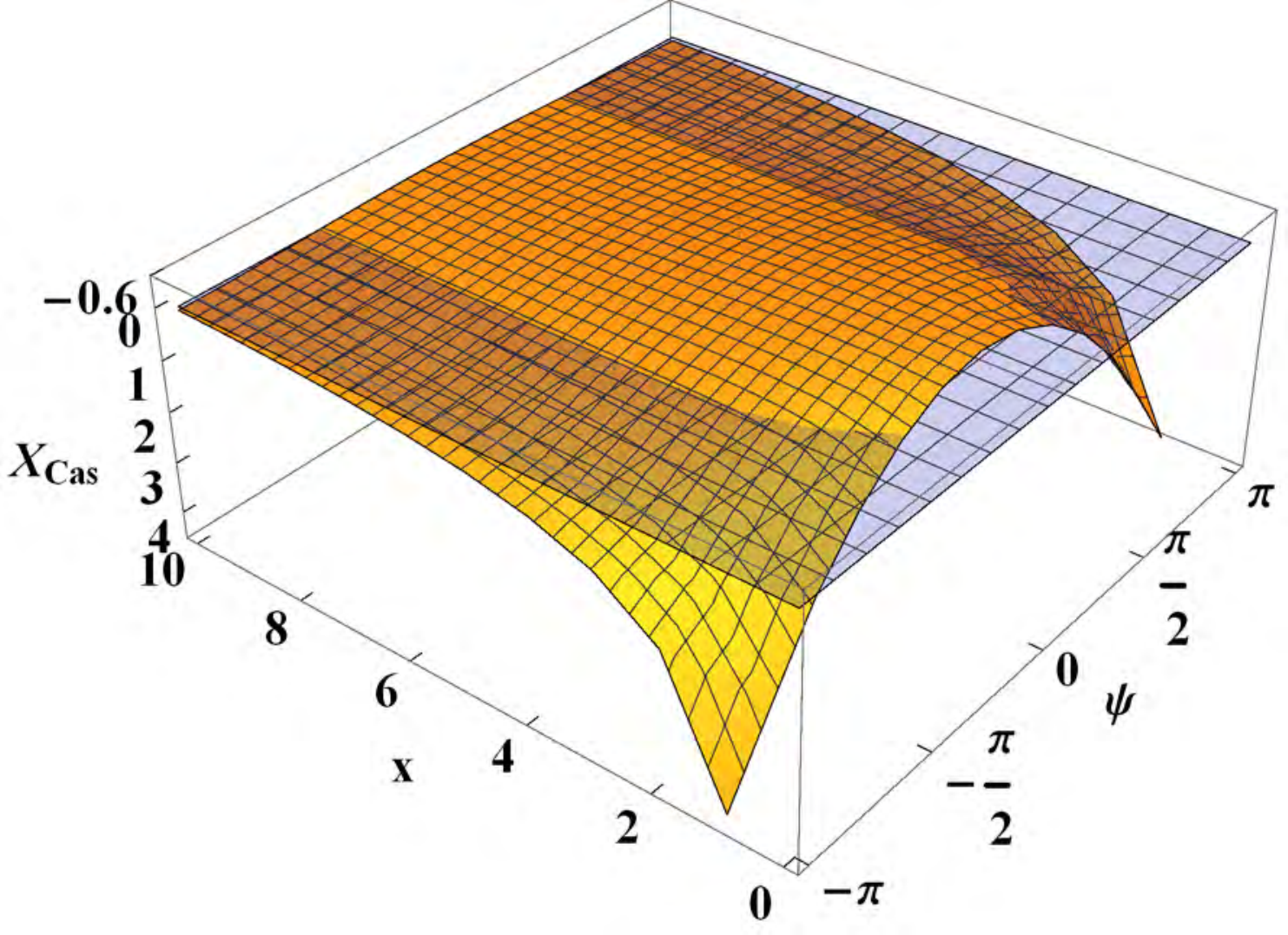}
\caption{(Color online) The surface of the scaling function $X_{\rm Cas}(\psi,x)$ of the XY model as a function of the scaling variables $x$ and $\psi$. The horizontal plane marks the $X_{\rm Cas}=0$ value.}
\label{Fig:1d_X_Cas_Surface}
\end{figure}

From \eq{eq:F_Cas_scaling_small_x} one can also derive an expression for the low $T$ behavior of the system that retains the dependence on $H_1$ and $H_2$. The result is
\begin{eqnarray}
\label{low_T_behavior}
\beta F_{\rm Cas}&=&-\frac{1}{2}\frac{1}{
   \left(J/H_1 + J/H_N+N-1\right)}\nonumber \\ && +\frac{1}{2}K\frac{\left(\psi _1-\psi
   _N\right)^2 }{
   \left(J/H_1+J/H_N+N-1\right)^2}.
\end{eqnarray}
This result can be also directly derived by realizing that the ground state of the system is a spin wave such that the end spins are twisted with respect to each other at angle $\psi=\psi_1-\psi_N$.

Equations \eqref{eq:F_Cas_scaling}, \eqref{eq:F_Cas_scaling_small_x},  \eqref{X_cas_ass} and \eqref{low_T_behavior} confirm the validity of the statements i)-iv) in the first part of this section. For example, \eq{eq:F_Cas_scaling} demonstrates that when $\psi=0$ the force is attractive for {\it any} value of the scaling variable $x$; \eq{X_cas_ass} then confirms this behavior for small and large values of the scaling variable $x$.

The behavior of the scaling function $X_{\rm Cas}(\psi,x)$ for different values of $\psi$ as a function of the scaling variable $x$ is shown in Fig. \ref{Fig:1d_X_Cas}. Fig. \ref{Fig:1d_X_Cas_Surface} shows a $3D$ plot of this function for $x\in[0,10]$ and $\psi\in[-\pi,\pi]$.

\subsection{The 1d Heisenberg model}
\label{sec:1dH_model}

The Hamiltonian of the system is again given by \eq{eq:def_1d_Ham} with the conditions that now the $N$ spins  ${\mathbf S}_i$, $i=1,\cdots N$, again arranged along a straight line, are three-dimensional vectors ${\mathbf S}_i \in \mathbb{Z}^3$, $i=1,\cdots,N$.
\begin{figure}[h]
\includegraphics[width=\columnwidth]{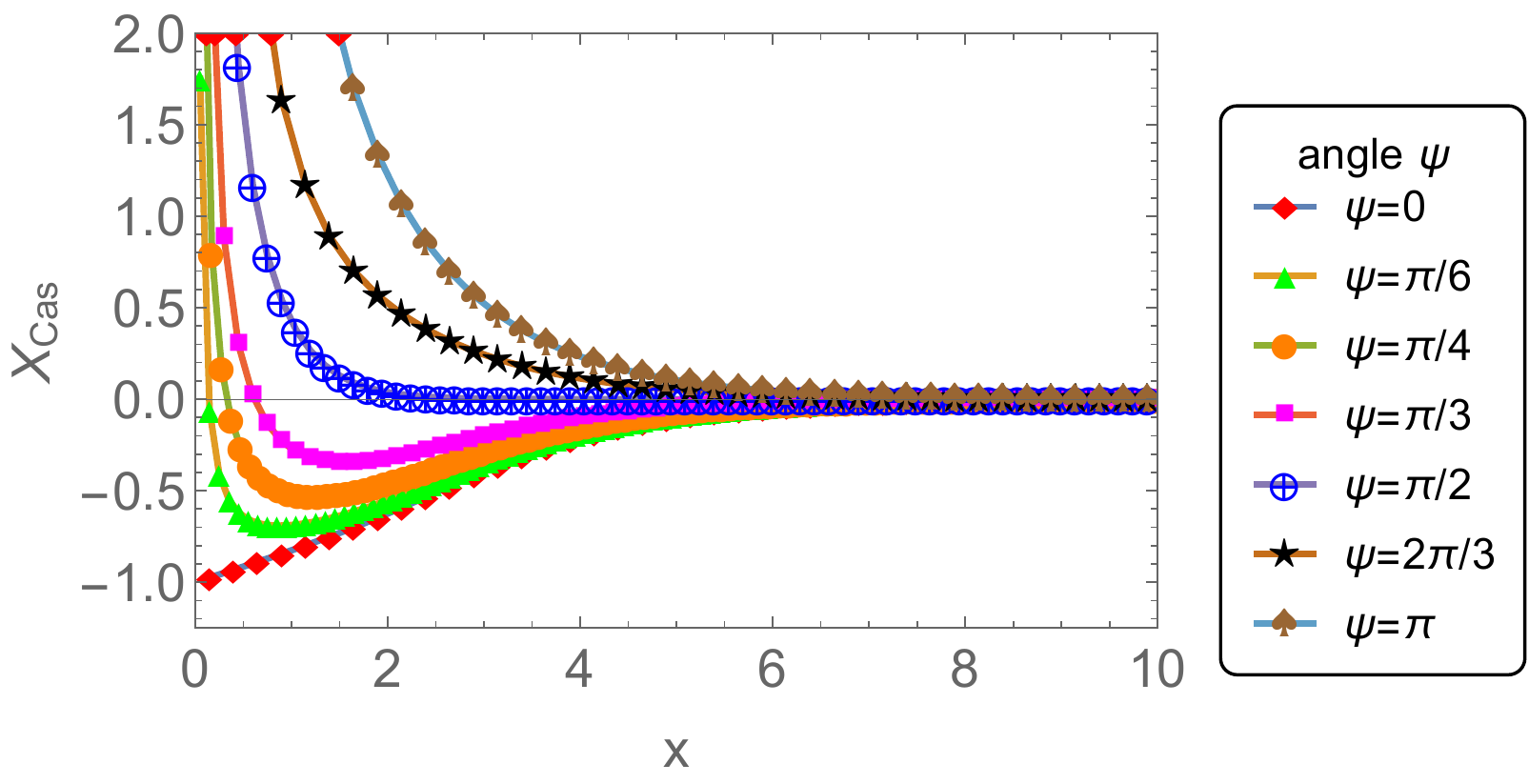}
\caption{(Color online) The scaling function $X_{\rm Cas}$ of the Heisenberg model as a function of the scaling variable $x$, see \eq{eq:def_FCas_Heis_scaling}, for different values of the phase change $\psi$.}
\label{Fig:1d_X_H_Cas}
\end{figure}

As shown in Appendix \ref{app:Heisenberg} the free energy of the system is given by the {\it exact} expression 
\begin{widetext}
\begin{eqnarray}
\label{eq:free_energy_calcul_final_result}
\exp \left(-\beta  F_N\right) &=& \left(\frac{\pi}{2K}\right)^{(N-1)/2}\frac{\pi}{2\sqrt{h_1h_N}} \sum_{n=0}^{\infty} (2n+1)P_n \left(\cos\psi_h\right) I_{n+1/2}(h_1) I_{n+1/2}(h_N) \left[I_{n+1/2}(K)\right]^{N-1}\\
&=& \frac{\sinh h_1}{h_1}\frac{\sinh h_N}{h_N}\left[\frac{\sinh K}{K}\right]^{N-1} \left\{1+ \sum_{n=1}^{\infty} (2n+1)P_n \left(\cos\psi_h\right) \frac{I_{n+1/2}(h_1)}{I_{1/2}(h_1)} \frac{I_{n+1/2}(h_N)}{I_{1/2}(h_N)} \left[\frac{I_{n+1/2}(K)}{I_{1/2}(K)} \right]^{N-1}\right\}, \nonumber
\end{eqnarray}
\end{widetext}
where $\psi_h$ is the angle between the vectors ${\mathbf H}_1$ and ${\mathbf H}_N$ and we have used that $I_{1/2}(x)=\sqrt{2/(\pi x)}\sinh(x)$. Here $I_{n+1/2}(z)$ is the modified Bessel function of the first kind of half-integer index, $P_n(x)$ is the Legendre polynomial of degree $n$ and $K$, $h_1$ and $h_N$ are defined in accord with \eq{eq:def_parameters_Heis}. 
\begin{equation}
\label{eq:def_parameters_Heis}
K\equiv \beta J, h_1\equiv \beta H_1, h_N\equiv\beta H_N. 
\end{equation}
When $h_1\to 0$ and $h_N\to 0$ the system considered becomes the one with Dirichlet boundary conditions, a case that was studied by M. E. Fisher in \cite{F64}. Taking into account that $I_{n+1/2}(x)=[2^{n+1/2}\Gamma(n+3/2)]^{-1}x^{n+1/2}+{\cal O}(x^{5/2+n})$ and that $P_0(x)=1$, one concludes that only the term with $n=0$ will contribute to the free energy in this case. One obtains 
\begin{eqnarray}
\label{eq:free_energy_calcul_final_result_Dirichlet_bc}
\exp \left(-\beta  F_N\right) &=& \left(\frac{\pi}{2K}\right)^{(N-1)/2}\left[I_{1/2}(K)\right]^{N-1}\\
&=&\left[\frac{\sinh K}{K}\right]^{N-1}. \nonumber
\end{eqnarray}
The last expression is precisely the result derived in \cite{F64}.

From \eq{eq:free_energy_calcul_final_result} one can easily derive the corresponding {\it exact} expression for the Casimir force for the one dimensional Heisenberg model. One has 
\begin{widetext}
\begin{equation}
\label{eq:def_FCas_Heis}
\beta  F_{\text{Cas}}=\frac{ \sum_{n=1}^{\infty} (2n+1)P_n \left(\cos\psi_h\right) \ln\left[\frac{I_{n+1/2}(K)}{I_{1/2}(K)}\right] \frac{I_{n+1/2}(h_1)}{I_{1/2}(h_1)} \frac{I_{n+1/2}(h_N)}{I_{1/2}(h_N)} \left[\frac{I_{n+1/2}(K)}{I_{1/2}(K)} \right]^{N-1}}{1+ \sum_{n=1}^{\infty} (2n+1)P_n \left(\cos\psi_h\right) \frac{I_{n+1/2}(h_1)}{I_{1/2}(h_1)} \frac{I_{n+1/2}(h_N)}{I_{1/2}(h_N)} \left[\frac{I_{n+1/2}(K)}{I_{1/2}(K)}\right]^{N-1}}.
\end{equation}
\end{widetext}

In the limit $T\to 0$ when $h_1\gg 1$, $h_N\gg 1$ and $K\gg 1$ from \eq{eq:as_form_SP} one obtains 
\begin{equation}
 \label{eq:F_Cas_scaling_Heis}
 \beta F_{\rm Cas}(x)=\frac{1}{N_{\rm eff}}X_{\rm Cas}(\psi_h,x,h_{\rm eff})
 \end{equation}
where the scaling variable $x$, as well as $h_{\rm eff}$, are as defined in \eq{eq:scaling_variables} while the scaling function $X_{\rm Cas}$ is 
\begin{widetext}
\begin{equation}
\label{eq:def_FCas_Heis_scaling}
X_{\rm Cas}(\psi_h,x,h_{\rm eff})=-\frac{1}{2}x\frac{ \sum_{n=1}^{\infty} n(n+1)(2n+1)P_n \left(\cos\psi_h\right)  \exp\left[-\frac{1}{2} n(n+1)\left(x+h_{\rm eff}^{-1}\right)\right]}{1+ \sum_{n=1}^{\infty}(2n+1)P_n \left(\cos\psi_h\right)  \exp\left[-\frac{1}{2} n(n+1)\left(x+h_{\rm eff}^{-1}\right)\right]}.
\end{equation}
\end{widetext}
As in the case of the $XY$ model, when \eq{eq:constraint} is fulfilled one can ignore $h_{\rm eff}$ in the above expression. If not stated otherwise we will always suppose this to be the case. Then the scaling function  $X_{\rm Cas}$ depends only on the scaling variable $x$ and the angle $\psi_h$ that parametrizes the boundary conditions on the system, exactly as set forth in \eq{eq:1dCas_gen}. The representation of $X_{\rm Cas}$ given by \eq{eq:def_FCas_Heis_scaling} is applicable for all values of $x$ except in the limit $x \ll 1$. Keeping in mind that $P_1(\cos \psi_h)=\cos\psi_h$, and in light of the fast decay off the terms in the sums in \eq{eq:def_FCas_Heis_ass}, it is clear that for those very small values of $x$ the sign of the force will be determined by the sign of $\cos \psi_h$. For the leading behavior of the Casimir force when $x\ll 1$ one obtains 
\begin{widetext}
\begin{equation}
\label{eq:def_FCas_Heis_ass}
X_{\rm Cas}(\psi_h,x,h_{\rm eff})=-1+
\frac{h_{\rm eff}^{-1}}{h_{\rm eff}^{-1}+x}
+
\frac{x (1-\cos\psi_h)}{\left(h_{\rm eff}^{-1}+x\right)^2}
+x\frac{
   \coth 
   \left(\frac{1}{h_{\rm eff}^{-1}+x}\right)
   -1}{\left(h_{\rm eff}^{-1}+x\right)^2},
\end{equation}
\end{widetext}
which follows from \eq{smallx_Heis}. 
One can also derive the first three terms in that expansion by considering the $N$ dependence of the ground energy of the 1d Heisenberg model, assuming it to be in the form of a spin wave. Explicitly, for the behavior of the Casimir force for $T\to 0$ from \eq{eq:def_FCas_Heis_ass} one obtains 
\begin{eqnarray}
\label{low_T_behavior_Heis}
\beta F_{\rm Cas}&=&-\frac{1}{
   \left(J/H_1 + J/H_N+N-1\right)}\nonumber \\ && +K \frac{1-\cos\psi_h}{
   \left(J/H_1+J/H_N+N-1\right)^2}.
\end{eqnarray}

\begin{figure}[h]
\includegraphics[width=\columnwidth]{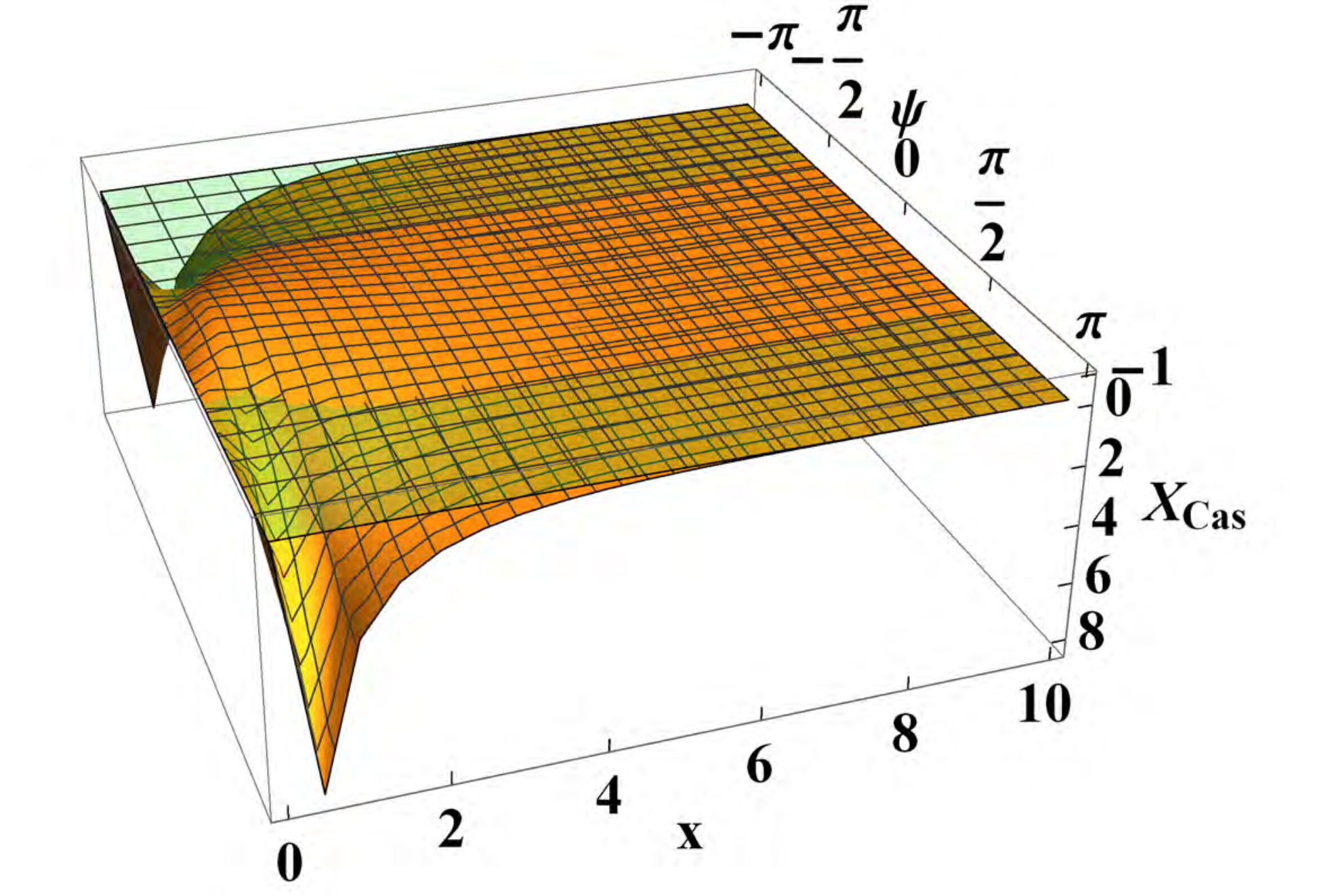}
\caption{(Color online) The surface of the scaling function $X_{\rm Cas}(\psi,x)$ of the Heisenberg model as a function of the scaling variables $x$ and $\psi$. The horizontal plane marks the $X_{\rm Cas}=0$ value.}
\label{Fig:1d_X_H_Cas_Surface}
\end{figure}

The behavior of the scaling function $X_{\rm Cas}(\psi,x)$ for different values of $\psi$ as a function of the scaling variable $x$ is shown in Fig. \ref{Fig:1d_X_H_Cas} while Fig. \ref{Fig:1d_X_H_Cas_Surface} shows a $3D$ plot of this function for $x\in[0,10]$ and $\psi\in[-\pi,\pi]$.  
Thus, for the overall behavior of the Casimir force as a function of $\psi_h$ one arrives at the same set of conclusions for the  Heisenberg model as for the $XY$ model as a function of $\psi$, as summarized in statements i)-v). 

\section{The 3d Gaussian model} \label{sec:Gaussian}

Here, we focus on a system with scalar spins. This means that, strictly speaking, there is no helicity. However, the surface fields that influence the order parameter will have sinusoidal variation along the film boundaries, conforming to the behavior of the individual components of a field that induces helical order in a multi-component system. We therefore expect that the results to be derived and discussed in this section will be germane to corresponding behavior in such a system. We consider a planar discrete system containing $L$ two-dimensional layers with a Hamiltonian
\begin{widetext}
\begin{eqnarray}
\label{eq:def_Ham_GM}
-\beta \mathcal{H}&=&\sum _{x=1}^M \sum _{y=1}^N \Bigg\{K^\| \sum _{z=1}^L S_{x,y,z} \left(S_{x+1,y,z}+S_{x,y+1,z}\right)+K^\perp\sum _{z=1}^{L-1} 
   S_{x,y,z} S_{x,y,z+1}+h_1 S_{x,y,1} \cos  \left(k_x x+k_y y\right) \nonumber\\
  &&  +h_L S_{x,y,L} \cos \left(k_x \left(x+\Delta _x\right)+k_y \left(y+\Delta
   _y\right)\right)-s \sum_{z=1}^L S_{x,y,z}^2\Bigg\}
\end{eqnarray}
\end{widetext}
which describes a system with short-ranged nearest neighbor interactions possessing chemically modulated bounding surfaces situated at $z=1$ and $z=L$. Here $h_1=\beta H_1$ and $h_L=\beta H_L$ are the external fields acting only on the boundaries of the system. In the specific example considered the modulation depends on the coordinates $x$ and $y$ in a wave-like way specified by the applied surface fields $h_1\cos\left(k_x x+k_y y\right)\equiv h_1 \cos({\bf k}.{\bf r})$ and $h_L \cos [k_x \left(x+\Delta_x\right)+k_y \left(y+\Delta_y\right)]\equiv h_L\cos({\bf k}.({\bf r}+{\bf \Delta}))$, the phases of which are thus shifted with respect to each other by $\Delta_x$ in $x$ direction and by $\Delta_y$ in $y$ direction. Here ${\bf r}=(x,y)$, ${\bf k}=(k_x,k_y)$ and ${\bf \Delta}=(\Delta_x,\Delta_y)$. Periodic boundary conditions are applied along the $x$ and $y$ axes, while missing neighbor (Dirichlet) boundary conditions are imposed in the $z$ direction. These boundary conditions are expressed as follows:
\begin{equation}
\label{bc_def_per}
S_{1,y,z}=S_{M+1,y,z}, \qquad S_{x,1,z}=S_{x,N+1,z}
\end{equation}
and
\begin{equation}
\label{bc_Dirichlet}
S_{x,y,0}=0 \qquad \text{and} \qquad S_{x,y,L+1}=0.
\end{equation}

Given those the boundary conditions, the Hamiltonian  in Eq. (\ref{eq:def_Ham_GM}) can be rewritten in the form 
\begin{eqnarray}
\label{eq:def_Ham_GM_final}
-\beta \mathcal{H}&=&\sum _{x=1}^M \sum _{y=1}^N \sum _{z=1}^L S_{x,y,z} \Bigg\{K^\| \left(S_{x+1,y,z}+S_{x,y+1,z}\right)  \nonumber \\ && + K^\perp
   S_{x,y,z+1} +\delta_{1,z} h_1 \cos  \left[{\bf k}.{\bf r} \right] \nonumber\\
  &&  + \delta_{L,z} h_L \cos \left[{\bf k}.({\bf r}+{\bf \Delta}) \right]-s\; S_{x,y,z}\Bigg\}.
\end{eqnarray}
Since we will be considering the limit $M,N\to\infty$ we can always take the wave vector components $k_x$ and $k_y$ to coincide with $(2\pi p)/M $ and $(2\pi q)/N$ for some $p=1,\cdots, M$ and $q=1,\cdots,N$, respectively. 
In Eqs. \eqref{eq:def_Ham_GM} and \eqref{eq:def_Ham_GM_final} one has 
\begin{equation}
\label{eq:inte}
K^{\|}=\beta J^{\|}, \qquad \mbox{and} \qquad K^{\perp}=\beta J^{\perp},
\end{equation}
where $J^{\|}$ and $J^{\perp}$ are the strengths of the coupling constants along and perpendicular to the $L$ layers of the system. The parameter $s>0$ on the right hand side of \eqref{eq:def_Ham_GM_final} is subjected to the constraint that it has a value that ensures the existence of the partition function of the system. It is easy to check that $2K^{\|}+K^{\perp}-s \equiv \beta (2J^{\|}+J^{\perp})-s=0$ determines the critical temperature $\beta_c$ of the bulk model, i.e., one has
\begin{equation}
\beta_c=s/(2J^{\|}+J^{\perp}).
\label{betac}
\end{equation}

For the model defined above the Casimir force acting on the bounding planes at $z=1$ and $z=L$ has both orthogonal, $\beta F^{(\perp)}_{\rm Cas}$, and lateral, $\beta F^{(\|,\alpha)}_{\rm Cas}$, $\alpha=x$ or $\alpha=y$, components, which can be written in the form 
\begin{equation}
\label{eq:gen_force}
\beta F^{(\cdots)}_{\rm Cas}=L^{-3}\left(\frac{J^\perp}{J^\|}\right) X^{(\cdots)}_{\rm Cas}(x_t,x_k,x_1,x_L),
\end{equation}
where $(\cdots)$ stands for either $(\perp)$ or $(\|,\alpha)$, with $\alpha=x$ or $\alpha=y$. Here 
\begin{equation}
\label{eq:field_scaling_def}
x_{1}=\sqrt{L K^\|}\frac{h_1}{K^\perp}, \qquad x_{L}=\sqrt{L K^\|}\frac{h_L}{K^\perp},
\end{equation}
are the field-dependent scaling  variables, $x_t$  is the temperature-dependent one with 
\begin{equation}
\label{eq:xt_and_xk}
x_t=L\sqrt{2\left(\frac{\beta_c}{\beta}-1\right)\left[ 2\frac{J^{\|}}{J^\perp}+1\right]}, \qquad x_k=\sqrt{\dfrac{J^{\|}}{J^{\perp}}}\; L k, 
\end{equation}
with $k=\sqrt{k_x^2+k_y^2}$ is the scaling variable related to the surface modulation. When $h_1={\cal O}(1)$ and $h_L={\cal O}(1)$ we will see that $F^{(\cdots)}_{\rm Cas}$ has a {\it field dependent contribution} which, in this regime, will provide the {\it leading} contribution to the force of the order of $L^{-2}$.

The Hamiltonian (\ref{eq:def_Ham_GM_final}) can be easily diagonalized in a standard way---see Appendix \ref{A:GM}. The resulting free energy of the system,  $F$, is  
\begin{equation}
\label{eq:fe_short}
F=\Delta F_0+\Delta F_h,
\end{equation}
where 
\begin{eqnarray}
\label{eq:free_energy_GM}
\lefteqn{-\beta \Delta F_0 = \frac{1}{2} M N L \ln\pi} \\
&& -\frac{1}{2}\sum _{l=1}^L \sum _{m=1}^M
   \sum _{n=1}^N \ln\left\{
   s-K^{\|}\left[\cos \left(\frac{2 \pi 
      m}{M}\right)+\cos \left(\frac{2 \pi 
         n}{N}\right)\right]\nonumber \right. \\
         &&\left.-K^{\perp} \cos\left(\frac{\pi 
            l}{L+1}\right)\right\} \nonumber
\end{eqnarray}
is the field independent part of the free energy and $\Delta F_h$,  the field dependent contribution, is 

i) when either $p\ne M$ or $q\ne N$:
\begin{eqnarray}
\label{eq:free_energy_h}
\lefteqn{-\beta \Delta F_h =  \frac{MN}{8(L+1)} \times} \\
            && \sum _{l=1}^L \frac{\sin^2\left(\frac{\pi l}{L+1}\right)\left[h_1^2+h_L^2-2 h_L h_1
               (-1)^l \cos(\mathbf{k.\Delta})\right]}{s-K^{\|}\left[\cos \left(\frac{2 \pi p
                  }{M} \right)+\cos \left(\frac{2 \pi q
                     }{N} \right)\right]-K^{\perp} \cos\left(\frac{\pi l}{L+1}\right)}, \nonumber
\end{eqnarray}
where $\mathbf{k}=(k_x=2 \pi  p /{M},k_y=2 \pi  q /{N})$, and  $\mathbf{\Delta}=(\Delta_x,\Delta_y)$, and 

ii) when $p=M$ and $q=N$:
\begin{eqnarray}
\label{eq:free_energy_h_cf}
\lefteqn{-\beta \Delta F_h =  \frac{MN}{2(L+1)} \times} \\
            && \sum _{l=1}^L \frac{\sin^2\left(\frac{\pi l}{L+1}\right)\left[h_1-h_L
               (-1)^l \cos \left(2 \pi (\Delta
               _x+\Delta
               _y)\right)\right]^2}{s-2K^{\|}-K^{\perp} \cos\left(\frac{\pi l}{L+1}\right)}. \nonumber
\end{eqnarray}
Note that there is a fundamental difference between the sub-cases in Eqs. \eqref{eq:free_energy_h} and \eqref{eq:free_energy_h_cf};  while in the first sub-case $i)$ the average field applied on the surfaces is zero when specially averaged, in the second sub-case  $ii)$ it is a constant. In the last sub-case one can think of $h_L$ as a constant field acting on the second surface being twisted in direction with respect to the constant field $h_1$ applied to  the first one with a twist governed by $\Delta_x$ and $\Delta_y$. 

Obviously  
\begin{eqnarray}
\lefteqn{s - K^{\|}\left[\cos \left(\frac{2 \pi 
   m}{M}\right)+\cos \left(\frac{2 \pi 
      n}{N}\right)\right] -
   K^{\perp} \cos\left(\frac{\pi 
   k}{L+1}\right)} \nonumber\\
   &&=\left(\beta_c/\beta-1\right)\left[ 2K^{\|}+K^\perp\right] +K^{\perp} \left[1-  \cos\left(\frac{\pi 
                     k}{L+1}\right)\right]\nonumber \\
         &&+K^{\|}\left[2-\cos \left(\frac{2 \pi 
               m}{M}\right)-\cos \left(\frac{2 \pi 
                  n}{N}\right)\right] >0
\end{eqnarray}
for $\beta<\beta_c$. The above implies that the statistical sum of the infinite system exists for all $\beta<\beta_c$. The statistical sum of the finite system exists, however, under the less demanding constraint that 
\begin{equation}
\label{eq:T_finite}
\left(\beta_c/\beta-1\right)\left[ 2J^{\|}+J^\perp\right] +J^{\perp} \left[1-  \cos\left(\frac{\pi 
            }{L+1}\right)\right]>0.
\end{equation}
In the remainder we will assume that the constraint given by  \eq{eq:T_finite} is fulfilled for all temperatures considered here. 

\begin{figure}[h]
\includegraphics[width=\columnwidth]{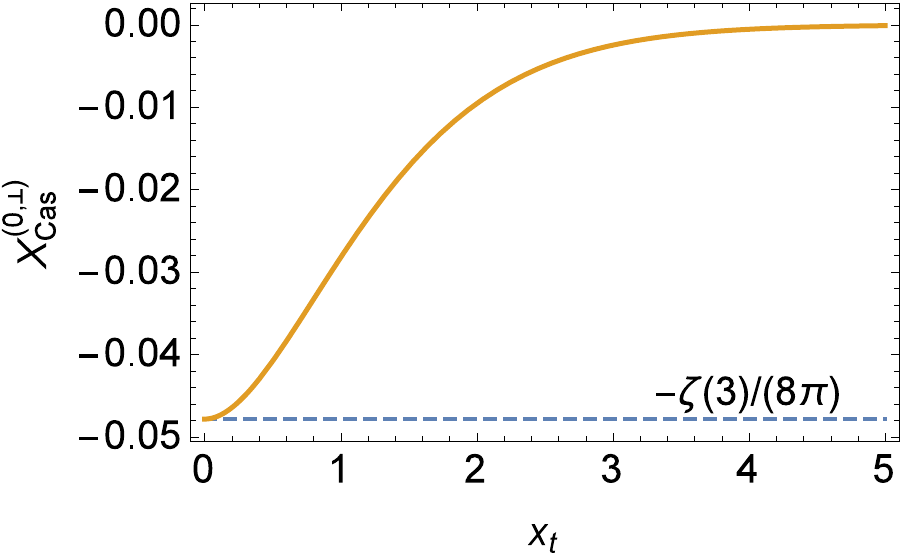}
\caption{(Color online) The scaling function $X^{(0,\perp)}_{\rm Cas}(x_t)$ as a function of the temperature dependent scaling variable $x_t$ The horizontal line marks the Casimir amplitude $X^{(0,\perp)}_{\rm Cas}(0)=-\zeta(3)/(8\pi)$.}
\label{Fig:3d_G_X_Cas_Zero_Field}
\end{figure}

For the  contribution of the field-independent term to the transverse Casimir force 
\begin{equation}
\label{eq:_def_no_field}
\beta\Delta F^{(0,\perp)}_{\rm Cas}=-\dfrac{\partial}{\partial L}(\beta\Delta f_0),
\end{equation}
with 
\begin{equation}
\label{eq:f_0}
\Delta f_0=\lim_{M, N \to \infty}
\dfrac{\Delta F_0}{MN},
\end{equation}
it is demonstrated in Appendix \ref{A:GM} that 
\begin{equation}
\label{eq:Cas_no_field}
\beta\Delta F^{(0,\perp)}_{\rm Cas}=-\frac{1}{2}\int _{-\pi }^{\pi
   }\int _{-\pi }^{\pi }\delta
   \left[\coth ((1+L) \delta)-1\right] \frac{d\theta
   _1d\theta _2}{ (2 \pi )^2},
\end{equation}
where $\delta=\delta\left(\theta_1,\theta_2|\beta_c/\beta,J^\|/J^\perp\right)$ is given by the expression
\begin{eqnarray}
\label{eq:def_delta}
\cosh\delta &=& 1+\left(\frac{\beta_c}{\beta}-1\right)\left(1+ 2\frac{J^{\|}}{J^\perp}\right)\\
&& +\frac{J^\|}{J^\perp}\left(2-\cos \theta_1-\cos \theta_2\right). \nonumber
\end{eqnarray}
The result in \eq{eq:Cas_no_field} is an {\it exact} expression for $\beta\Delta F^{(0,\perp)}_{\rm Cas}$; no approximations have been made. Since $\coth(x)>1$ for $x>0$ one immediately concludes that $\Delta F^{(0,\perp)}_{\rm Cas}<0$, i.e., it is an {\it attractive} force, for {\it all} values of $L$. In order to obtain scaling and, thus, the scaling form of $\Delta F^{(0,\perp)}_{\rm Cas}$ we have to consider the regime $L\gg 1$. 
Obviously, then Casimir force will be exponentially small if $\delta$ is finite. For the scaling behavior of the force---see Appendix \ref{A:GM}---one obtains 
\begin{equation}
\label{eq:F_Cas_no_field}
\beta\Delta F^{(0,\perp)}_{\rm Cas}=L^{-3}\left(\frac{J^\perp}{J^\|}\right) X^{(0,\perp)}_{\rm Cas}(x_t)
\end{equation}
where $X^{(0,\perp)}_{\rm Cas}(x_t)$ is the universal scaling function 
\begin{eqnarray}
\label{eq:X_Cas_no_field_sf}
X^{(0,\perp)}_{\rm Cas}(x_t)&=&-\frac{1}{8 \pi}\Bigg\{
      \text{Li}_3\left(e^{-2
      x_t}\right)+2 x_t
   \text{Li}_2\left(e^{-2
   x_t}\right) \nonumber \\
   && -2 x_t^2 \ln \left(1-e^{-2
   x_t}\right)\Bigg\}.
\end{eqnarray}
and the scaling variable $x_t$ is
 \begin{equation}
\label{eq:delta_xt}
x_t=L \sqrt{2\left(\frac{\beta_c}{\beta}-1\right)\left(1+ 2\frac{J^{\|}}{J^\perp}\right)},
\end{equation}
in accord with  \eq{delta_xt}. It is easy to show that $X^{(0,\perp)}_{\rm Cas}(x_t)$ is a {\it monotonically increasing} function of $x_t$. The behavior of $X^{(0,\perp)}_{\rm Cas}(x_t)$ is visualized in Fig. 
\ref{Fig:3d_G_X_Cas_Zero_Field}

At the critical point one has $x_t=0$ and then one immediately obtains the well known Casimir amplitude for the Gaussian model under Dirichlet boundary condition 
\begin{equation}
\label{eq:F_Cas_no_field_ampl}
X^{(0,\perp)}_{\rm Cas}(x_t=0)=-\frac{\zeta(3)}{8 \pi}.
\end{equation}

It is easy to show that
\begin{equation}
\label{eq:asX0_Cas}
X^{(0,\perp)}_{\rm Cas}\simeq 
\left\{\begin{array}{lcr}
-\frac{1}{8\pi} \exp(-2 x_t) \left[1+2 x_t
   \left(1+x_t\right)\right],& x_t\gg 1&\\
   && \\
   -\frac{1}{8 \pi
      }\zeta (3)+\frac{1}{48 \pi} x_t^2 \left(6-4
      x_t+x_t^2\right),& x_t\to 0.
\end{array} \right.
\end{equation}

For the field component of the transverse Casimir force 
\begin{equation}
\label{eq:_def_field}
\beta\Delta F^{(h,\perp)}_{\rm Cas}=-\dfrac{\partial}{\partial L}(\beta\Delta f_h)
\end{equation}
where 
\begin{equation}
\label{eq:f_h}
\Delta f_h=\lim_{M, N \to \infty}
\dfrac{\Delta F_h}{MN}
\end{equation}
one derives, see Eqs.  \eqref{eq:delta_fh_final_pq} and \eqref{eq:delta_fh_final} in Appendix  \ref{A:GM}: 

{\it i)} if $p\ne  M$ or $q\ne N$:
\begin{eqnarray}
\label{eq:Casimir_tr_pq}
\lefteqn{\beta \Delta F^{(h,\perp)}_{\rm Cas}=\frac{\lambda\sinh (\lambda)}{32 K^{\perp}}}\\
&&\times\left\{  \left[h_1^2+h_L^2-2 h_L h_1
                            \cos(\mathbf{k.\Delta})\right]^2 \text{csch}^2\left[\frac{1+L}{2} \lambda\right] \right.\nonumber\\
&&\left.-\left[h_1^2+h_L^2+2 h_L h_1
                            \cos(\mathbf{k.\Delta})\right]^2 \text{sech}^2\left[\frac{1+L}{2} 
   \lambda\right]\right\}. \nonumber
\end{eqnarray}
and 

{\it ii)} if $p=M$ and $q=N$ 
\begin{eqnarray}
\label{eq:Casimir_tr}
\lefteqn{ \beta \Delta F^{(h,\perp)}_{\rm Cas}=\frac{\lambda\sinh (\lambda)}{32 K^{\perp}} }\\
&&\times\left\{  \left[h_1-h_L
                                       \cos 2 \pi  ( \Delta
                                       _x+ \Delta
                                       _y)\right]^2 \text{csch}^2\left[\frac{1+L}{2} \lambda\right] \right.\nonumber\\
&&\left.-\left[h_1+h_L
                                       \cos 2 \pi  ( \Delta
                                       _x+ \Delta
                                       _y)\right]^2 \text{sech}^2\left[\frac{1+L}{2} 
   \lambda\right]\right\}. \nonumber
\end{eqnarray}
Here we have introduced the helpful notation 
\begin{equation}
\label{eq:x_def_1}
\cosh \lambda=\Lambda
\end{equation}
for the case when $\Lambda\ge 1$ and 
\begin{equation}
\label{eq:x_def_2}
\cos \lambda=\Lambda
\end{equation}
in the opposite case when $\Lambda\le 1$. Note that
\begin{itemize}
	\item 
 when $h_1={\cal O}(1)$, $h_L={\cal O}(1)$ and 
\begin{equation}
\label{eq:w_def}
w=L\lambda/2
\end{equation}
is such that $w={\cal O}(1)$, {\it the Casimir force is of the order of }${\cal O}(L^{-2})$ { despite} the fact that the system is at a temperature {\it above} the bulk critical one. 

\item If $h_1$ and $h_L$ are such that  the field-dependent scaling  variables $x_1={\cal O}(1)$ and $x_L={\cal O}(1)$, see \eq{eq:field_scaling_def}, then, 
in terms of $w$, the Casimir force $\beta\Delta F^{(h,\perp)}_{\rm Cas}$ reads
\begin{equation}
\label{eq:sc_funct_field}
\beta\Delta F^{(h,\perp)}_{\rm Cas}=L^{-3}\left(\frac{J^\perp}{J^\|}\right) X^{(h,\perp)}_{\rm Cas}(w,x_1,x_L)
\end{equation}
where the scaling function $X^{(h,\perp)}_{\rm Cas}(w,x_1,x_L)$ is

{\it i)} if $p\ne  M$ or $q\ne N$:
\begin{eqnarray}
\label{eq:scaling_h_pq}
\lefteqn{X^{(h,\perp)}_{\rm Cas}(w,x_1,x_L)=\frac{1}{8} w^2}\\
&& \times \left\{[x_1^2+x_L^2-2x_1x_L \cos\left(\mathbf{k.\Delta}\right)] \text{csch}^2 w  \right. \nonumber \\
&&\left. -  [x_1^2+x_L^2+2x_1 x_L \cos\left(\mathbf{k.\Delta}\right)] \text{sech}^2 w \right\}, \nonumber
\end{eqnarray}
and 

{\it ii)} if $p=M$ and $q=N$ 
\begin{eqnarray}
\label{eq:scaling_h}
\lefteqn{X^{(h,\perp)}_{\rm Cas}(w,x_1,x_L) = \frac{1}{8} w^2}\\
&& \times \left\{[x_1-x_L \cos 2 \pi  ( \Delta
                                       _x+ \Delta
                                       _y)]^2 \text{csch}^2 w  \right. \nonumber \\
&&\left. -  [x_1+x_L \cos 2 \pi  ( \Delta
                                       _x+ \Delta
                                       _y)]^2 \text{sech}^2 w \right\}. \nonumber
\end{eqnarray}
The latter expression implies  that in the regime considered here the field-dependent part of the force if of order of $L^{-3}$, as it is the field-independent part of it. 
\end{itemize}

The asymptotic behavior of $\Delta F^{(h,\perp)}_{\rm Cas}$ for $w\gg 1$ can be easily obtained from Eqs. \eqref{eq:large_x_pq} and \eqref{eq:large_x}. The result is 
\begin{eqnarray}
\label{eq:scaling_h_as}
\lefteqn{\beta\Delta F^{(h,\perp)}_{\rm Cas}\simeq -\dfrac{2w^2}{K^{\perp}L^2} e^{-2w}h_1 h_L} \nonumber \\
&&\times \left\{
\begin{array}{ll}
\cos\left(\mathbf{k.\Delta}\right), & p\ne  M \quad \mbox{or} \quad q\ne N,
\\
\cos 2 \pi (\Delta_x+ \Delta_y), & p=M, q=N.
\end{array} \right.
\end{eqnarray}
which implies that in this limit the transverse component of the force is exponentially small in $L$ and attractive  {\it or } repulsive depending on the product $h_1 h_L \cos[\mathbf{k.\Delta}]$ or $h_1 h_L \cos 2 \pi (\Delta_x+ \Delta_y)$.

For the  field contribution to the longitudinal component of the Casimir force along the $\alpha$ axis, where $\alpha=x,y$, one has
\begin{equation}
\label{eq:_def_field_x}
\beta \Delta F^{(h,\alpha)}_{\rm Cas}(L)=-\dfrac{\partial}{\partial \Delta_\alpha}\Delta f_h.
\end{equation}
Thus, from Eqs. \eqref{eq:delta_fh_final_pq} and 
\eqref{eq:delta_fh_final}
one derives

{\it i)} if $p\ne  M$ or $q\ne N$:
\begin{equation}
\label{eq:Casimir_longit_pq}
\beta\Delta F^{(h,\alpha)}_{\rm Cas}(L)=-\frac{h_1 h_L }{4 K^{\perp}} k_\alpha \sin (\mathbf{k.\Delta}) \frac{\sinh(\lambda)}{\sinh[\lambda(L+1)]} 
\end{equation}
and 

{\it ii)} if $p=M$ and $q=N$ 
\begin{eqnarray}
\label{eq:Casimir_longit}
\lefteqn{\beta\Delta F^{(h,\alpha)}_{\rm Cas}(L) = -\frac{ \pi \sin [2 \pi (\Delta_x+ \Delta_y)]}{2 K^{\perp}} h_L} \\
&& \times \Bigg\{ h_1 \frac{\sinh(\lambda)}{\sinh[(L+1)\lambda]} \nonumber\\
&& + h_L \cos[2 \pi (\Delta_x+ \Delta_y)]  \left[\Lambda-\frac{\sinh(\lambda)}{\tanh (L+1)\lambda}\right]\Bigg\}. \nonumber
\end{eqnarray}

When $L\lambda\gg 1$ the above simplifies to 

{\it i)} if $p\ne  M$ or $q\ne N$:
\begin{equation}
\label{eq:scaling_h_as_x_pq}
\beta\Delta F^{(h,\alpha)}_{\rm Cas}(L) \simeq -\dfrac{k_\alpha}{2 K^{\perp}}  \sinh[\lambda] e^{-(L+1)\lambda} h_1 h_L \sin\left(\mathbf{k.\Delta}\right)
\end{equation}
and 

{\it ii)} if $p=M$ and $q=N$ 
\begin{eqnarray}
\label{eq:scaling_h_as_x}
\lefteqn{\beta\Delta F^{(h,\alpha)}_{\rm Cas}(L) \simeq } \\
&&  -\frac{ \pi h_L^2}{4 K^{\perp}} \sin [4 \pi (\Delta_x+ \Delta_y)] \left\lbrace \Lambda-\sinh[\lambda]\right\rbrace  \nonumber \\
&& -\dfrac{\pi}{K^{\perp}}  \sinh[\lambda] e^{-(L+1)\lambda} h_1 h_L \sin [2 \pi (\Delta_x+ \Delta_y)]. \nonumber
\end{eqnarray}
Note that in the first sub-case the $L\gg 1$ limit of the lateral force is zero, in the second sub-case, when the average value of the external field on the upper surface is not zero the lateral force tends to a finite, well defined limit which is proportional to the surface area of the system. Obviously, this force has the meaning of a local purely surface force.

Subtracting from  $\Delta F^{(h,\alpha)}_{\rm Cas}$ its $L$-independent part we obtain the lateral force that will act on the upper surface due to the presence of the lower one if we act in lateral direction on the upper one. In the case  $p=M$ and $q=N$  one obtains 
\begin{eqnarray}
\label{eq:F_long}
\lefteqn{\beta\delta F^{(h,\alpha)}_{\rm Cas}(L) \equiv  \beta\left[\Delta F^{(h,\alpha)}_{\rm Cas}(L)-\lim_{L\to\infty}\Delta F^{(h,\alpha)}_{\rm Cas}(L)\right]}\nonumber\\
&& =-\frac{ \pi h_L }{2 K^{\perp}} \sin [2 \pi (\Delta_x+ \Delta_y)] \sinh(\lambda)  \Bigg\{ h_1/ \sinh[(L+1)\lambda] \nonumber \\
&& + h_L \cos[2 \pi (\Delta_x+ \Delta_y)] [1-\coth (L+1)\lambda]\Bigg\}. 
\end{eqnarray}
In the other sub-case when $p\ne  M$ or $q\ne N$ one has that $\beta\delta F^{(h,\alpha)}_{\rm Cas}(L)\equiv \beta\Delta F^{(h,\alpha)}_{\rm Cas}(L)$. 

In scaling variables for $\beta\delta F^{(h,\alpha)}_{\rm Cas}(L)$ one has 
\begin{equation}
\label{eq:sc_funct_field_long}
\beta\delta F^{(h,\alpha)}_{\rm Cas}(L)=L^{-3}\left(\frac{J^\perp}{J^\|}\right) X^{(h,\alpha)}_{\rm Cas}(w,x_1,x_L),
\end{equation}
where 

{\it i)} if $p\ne  M$ or $q\ne N$:
\begin{equation}
\label{eq:Casimir_longit_pq_scaling}
X^{(h,\alpha)}_{\rm Cas}=-\pi x_1 x_L \, p_\alpha \sin (\mathbf{k.\Delta}) \frac{\omega}{\sinh[2\omega]}, 
\end{equation}
where $p_\alpha=p$ for $\alpha=x$, and $p_\alpha=q$ for $\alpha=y$. 

{\it ii)} if $p=M$ and $q=N$: 
\begin{eqnarray}
\label{eq:F_long_scaling}
\lefteqn{X^{(h,\alpha)}_{\rm Cas}=-\pi x_L \omega \sin [2 \pi (\Delta_x+ \Delta_y)] }  \\
&& \times \Bigg\{ x_1/ \sinh[2\omega] + x_L \cos[2 \pi (\Delta_x+ \Delta_y)] [1-\coth 2\omega]\Bigg\}. \nonumber 
\end{eqnarray}
\eq{eq:sc_funct_field_long} implies that in the scaling regime the longitudinal Casimir force is of the same order of magnitude as the orthogonal component of the force.

Let us now clarify the physical meaning of the regimes $\omega={\cal O}(1)$ and  $\omega\gg 1$ in terms of the temperature $T$. Taking into account \eq{eq:Lambda_def} one has
\begin{equation}
\label{eq:Lambda_def_delta}
\Lambda = 1+\left(\frac{\beta_c}{\beta}-1\right)\left[ 2\frac{J^{\|}}{J^\perp}+1\right]+2\dfrac{J^{\|}}{J^{\perp}}\left[\sin^2 \frac{k_x}{2} +\sin^2 \frac{k_y}{2}\right],
\end{equation}
where $k_x=2 \pi p/{M}$, $k_y=2 \pi q/{N}$, as well as all the other terms in the sum determining $\Lambda$ are dimensionless. We again have to consider two sub-cases:

{\it i)} if $p\ne  M$ or $q\ne N$.

In this case, in order to have $\lambda$ small, one needs to have $\beta/\beta_c\to 1$, and $k_\alpha\to 0$, $\alpha=x,y$. Under this conditions one has
\begin{equation}
\label{eq:lambda_eq}
\lambda \simeq \sqrt{2\left(\frac{\beta_c}{\beta}-1\right)\left[ 2\frac{J^{\|}}{J^\perp}+1\right]+\dfrac{J^{\|}}{J^{\perp}}\left[k_x^2 +k_y^2\right]}.
\end{equation}
Then
\begin{equation}
\label{eq:omega_eq}
\omega=\frac{1}{2}\sqrt{x_t^2+x_k^2},
\end{equation}
where $x_t$ and $x_k$ are defined in \eq{eq:xt_and_xk}. From \eq{eq:omega_eq} it is clear that in order to have $\omega = {\cal O}(1)$ one needs to have simultaneously $x_t={\cal O}(1)$ and $x_k={\cal O}(1)$. Taking into account that $\nu=1/2$ for the Gaussian model, one has that $x_t^2$ is in its expected form $a_t t L^{1/\nu}$, with $t=(T-T_c)/T_c$. The condition $x_k={\cal O}(1)$ implies that in order to encounter the regime $\omega = {\cal O}(1)$ one needs to have a modulation with a wave vector $k\lesssim L^{-1}$ which includes, e.g., the $k=0$ case. If $x_k\gg 1$ one will have, even at the critical point $\beta=\beta_c$ that $\omega\gg 1$ and, according to \eq{eq:scaling_h_as}, that the field contributions into the Casimir force will be exponentially small then. 

{\it ii)} if $p=M$ and $q=N$.

As it is clear from \eq{eq:Lambda_def_delta}, this sub-case reduces to the previously considered one with $k_x=k_y=0$. The last implies that, then, $\omega=x_t/2$. 

When $\omega = {\cal O}(1)$, from Eqs. \eqref{eq:Casimir_tr_pq} and \eqref{eq:Casimir_tr} with $h_1 = {\cal O}(1)$ and $h_L = {\cal O}(1)$ one has that   $\Delta F^{(h,\perp)}_{\rm Cas} = {\cal O}(L^{-2})$, i.e., the longitudinal force in this case is in an order of magnitude {\it larger} in $L$ than the usual transverse Casimir force, which is of the order of ${\cal O}(L^{-3})$.

\begin{figure}[h]
\includegraphics[width=\columnwidth]{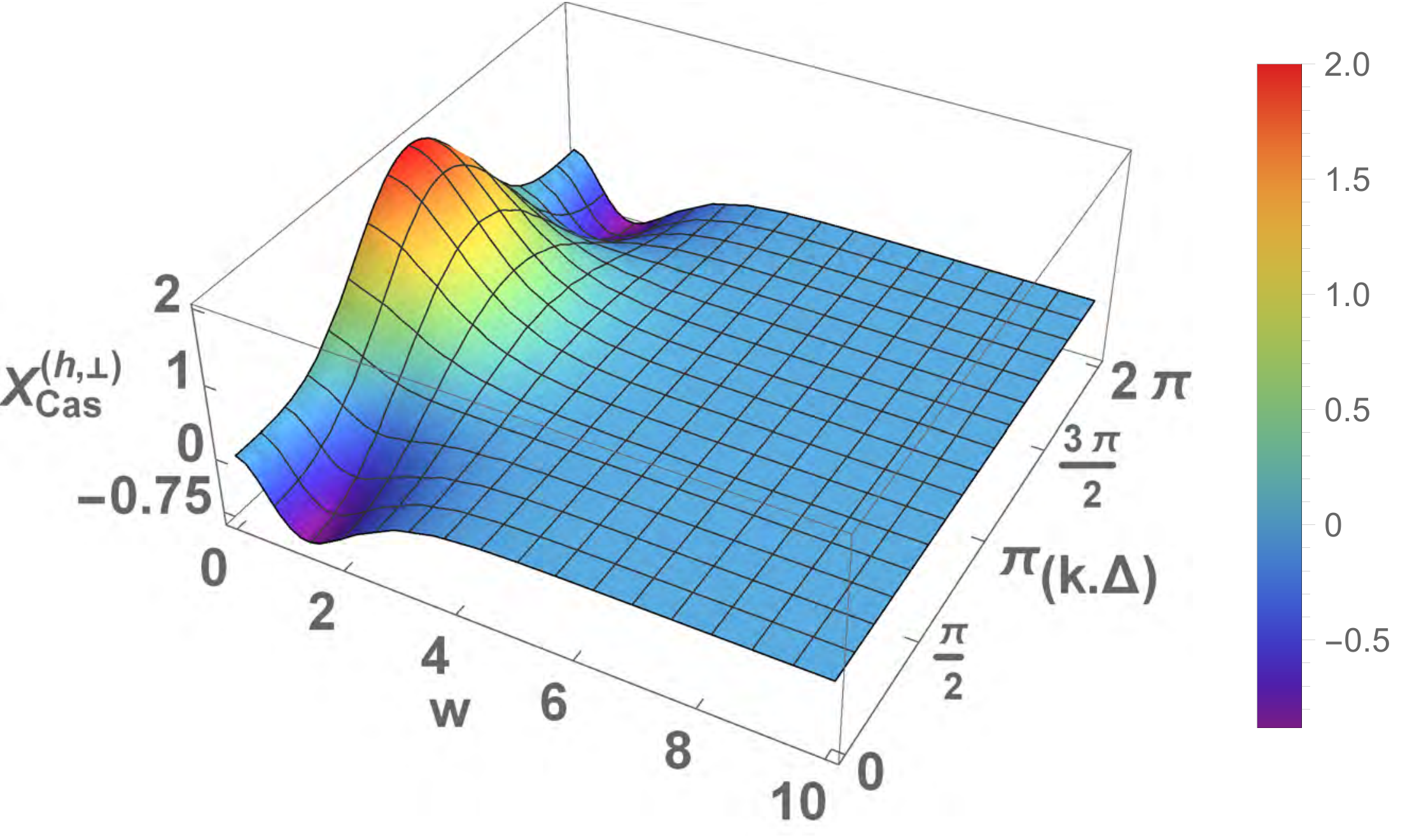}
\caption{(Color online) The scaling function $X^{(h,\perp)}_{\rm Cas}(w,x_1,x_L)$, see \eq{eq:scaling_h_pq}, as a function of $w\in (0,10]$ and $\left(\mathbf{k.\Delta}\right)\in [0,2\pi]$ for $x_1=x_L=1$. As wee see, $X^{(h,\perp)}_{\rm Cas}(w,x_1,x_L)$ can be both positive and negative, depending on the values of its arguments.}
\label{Fig:3D_G_h1_eq_hL_Legend}
\end{figure}

\begin{figure}[h]
\includegraphics[width=\columnwidth]{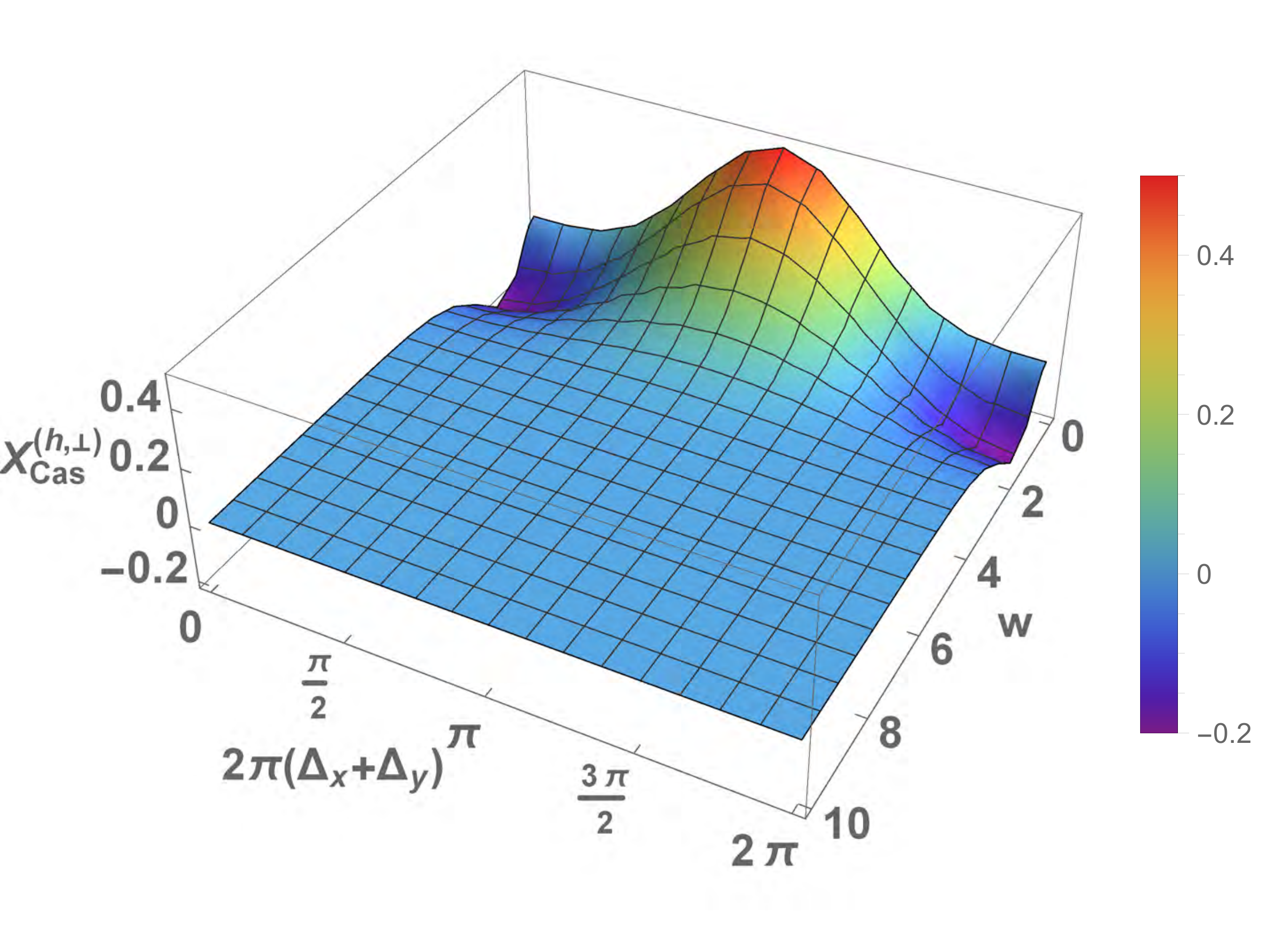}
\caption{(Color online) The scaling function $X^{(h,\perp)}_{\rm Cas}(w,x_1,x_L)$, see \eq{eq:scaling_h}, as a function of $w\in(0,10]$ and $\Delta_x+ \Delta_y\in [0,1]$ for $x_1=x_L=1$. As wee see, also in this case  $X^{(h,\perp)}_{\rm Cas}(w,x_1,x_L)$ can be both positive and negative, depending on the values of its arguments. Let us remind that in this sub-case $\omega=x_t/2$.  }
\label{Fig:3D_G_h1_eq_hL_Legend_MN}
\end{figure}

The behavior of the function $X^{(h,\perp)}_{\rm Cas}(w,x_1,x_L)$ is visualized in Fig. \ref{Fig:3D_G_h1_eq_hL_Legend} if {\it i)}  $p\ne  M$ or $q\ne N$ and in Fig. \ref{Fig:3D_G_h1_eq_hL_Legend_MN}
if {\it ii)} $p=M$ and $q=N$ . 

We observe, inspecting the legends, that the maximal values of the function $X^{(h,\perp)}_{\rm Cas}(w,x_1,x_L)$ are in this case smaller than in previous case shown in Fig. \ref{Fig:3D_G_h1_eq_hL_Legend}. 

Let us turn now to the behavior of the total orthogonal Casimir force $F^{(\perp)}_{\rm Cas}$. From Eqs. \eqref{eq:fe_short}, \eqref{eq:_def_no_field}, \eqref{eq:f_0}, \eqref{eq:F_Cas_no_field}, \eqref{eq:_def_field}  and  \eqref{eq:sc_funct_field} one has 
\begin{equation}
\label{eq:ort_force}
F^{(\perp)}_{\rm Cas}\equiv \Delta F^{(0,\perp)}_{\rm Cas}+\Delta F^{(h,\perp)}_{\rm Cas}
\end{equation}
and 
\begin{equation}
\label{eq:ort_force_sf}
\beta F^{(\perp)}_{\rm Cas}=L^{-3}\left(\frac{J^\perp}{J^\|}\right) X^{(\perp)}_{\rm Cas}(x_t,x_k,x_1,x_L).
\end{equation}

\begin{figure}[h]
\centering
\includegraphics[width=\columnwidth]{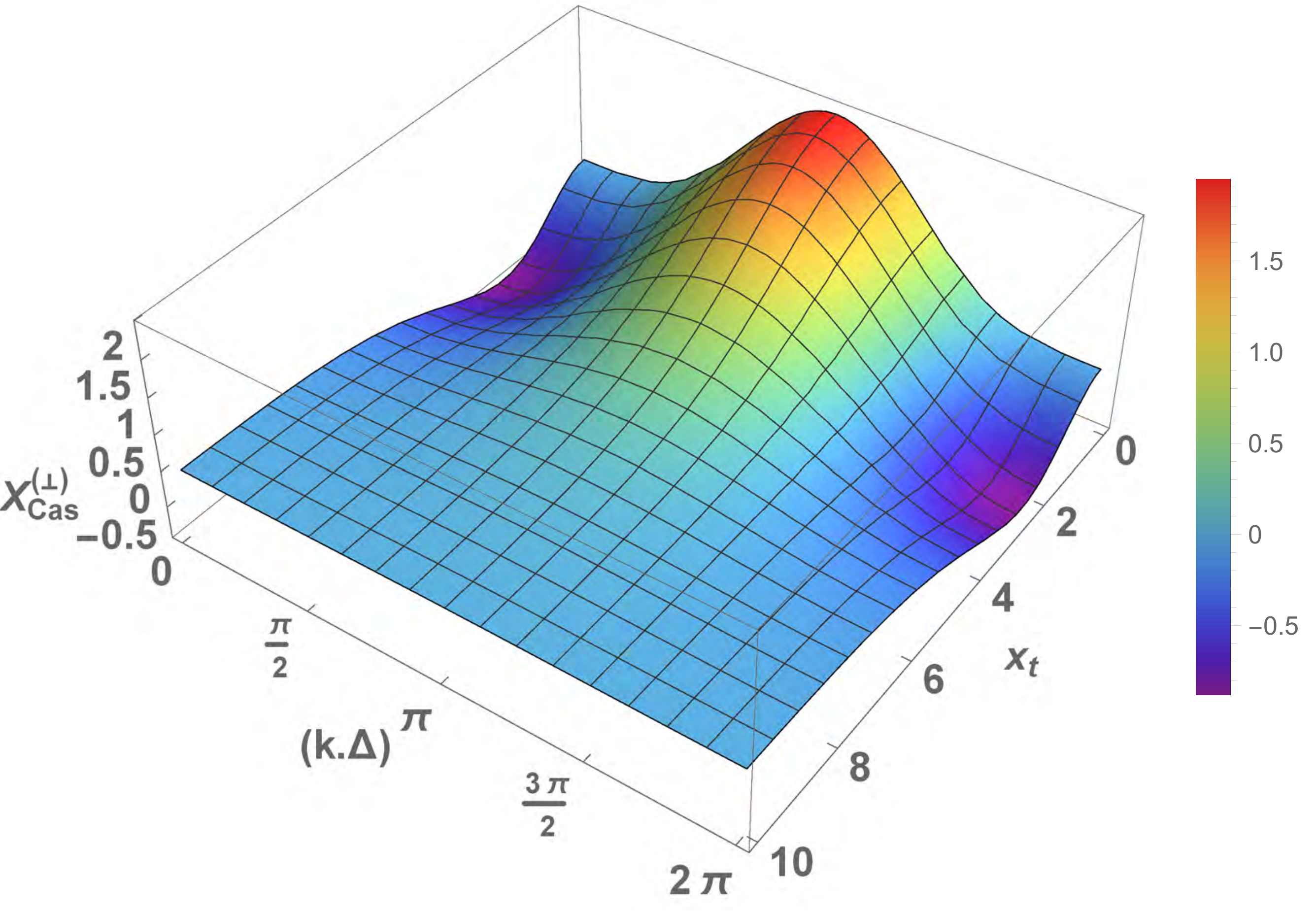}
\caption{(Color online) The scaling function 
$X^{(\perp)}_{\rm Cas}(x_t,x_k,x_1,x_L)$
 as a function of $x_t\in(0,10]$ and $\mathbf{k.\Delta}\in [0,2\pi]$ for $x_k=0.1$, $x_1=x_L=1$. As wee see, $X^{(\perp)}_{\rm Cas}$ can be both positive and negative, depending on the values of its arguments.}
\label{Fig:3D_G_h1_eq_hL_Legend_total_force}
\end{figure}

\begin{figure}[h]
\includegraphics[width=\columnwidth]{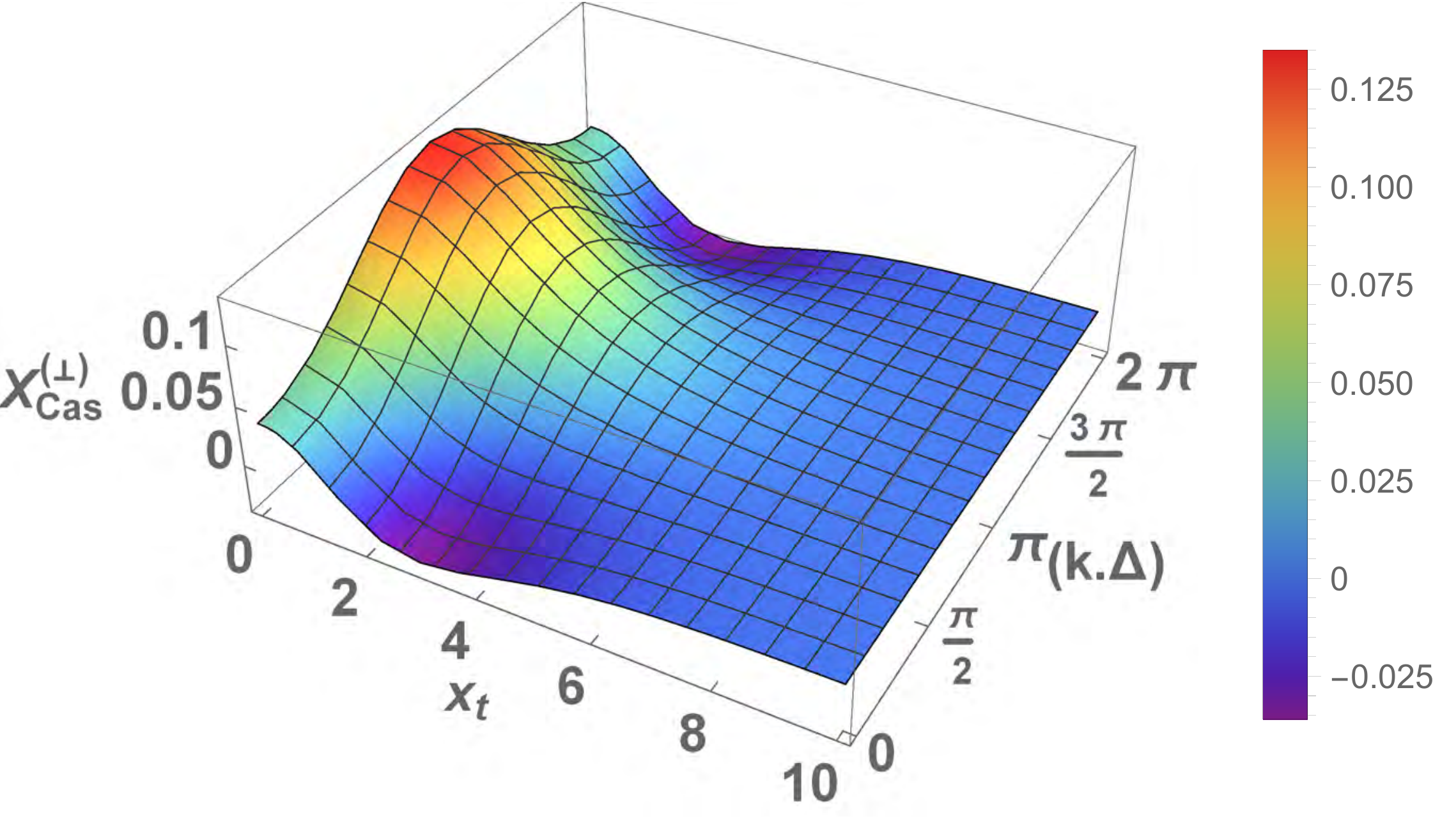}
\caption{(Color online) The scaling function 
$X^{(\perp)}_{\rm Cas}(x_t,x_k,x_1,x_L)$
 as a function of $x_t\in(0,10]$ and $\mathbf{k.\Delta}\in [0,2\pi]$ for $x_k=0.1$, $x_1=10 x_L=1$. As wee see, the scaling function in that case is predominantly positive.}
\label{Fig:3D_G_h1_not_eq_hL_Legend_total_force_1}
\end{figure}

\begin{figure}[h]
\includegraphics[width=\columnwidth]{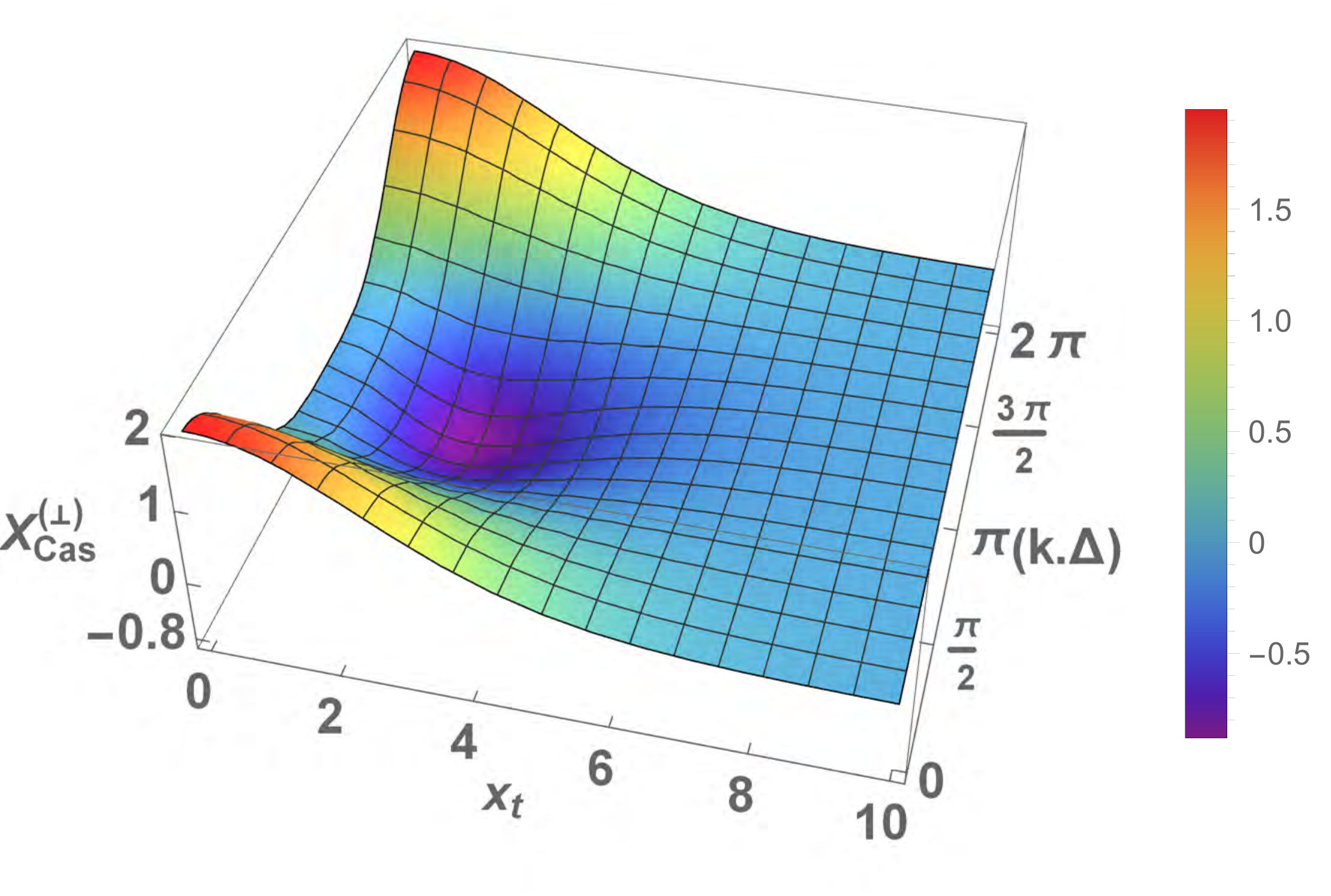}
\caption{(Color online) The scaling function 
$X^{(\perp)}_{\rm Cas}(x_t,x_k,x_1,x_L)$
 as a function of  $x_t\in(0,10]$ and $\mathbf{k.\Delta}\in [0,2\pi]$ for $x_k=0.1$, $ x_1=-x_L=1$. As wee see, the scaling function in that case can be both positive and negative, depending on the values of its arguments.}
\label{Fig:3D_G_h1_not_eq_hL_Legend_total_force_2}
\end{figure}

\begin{figure}[h]
\centering
\includegraphics[width=\columnwidth]{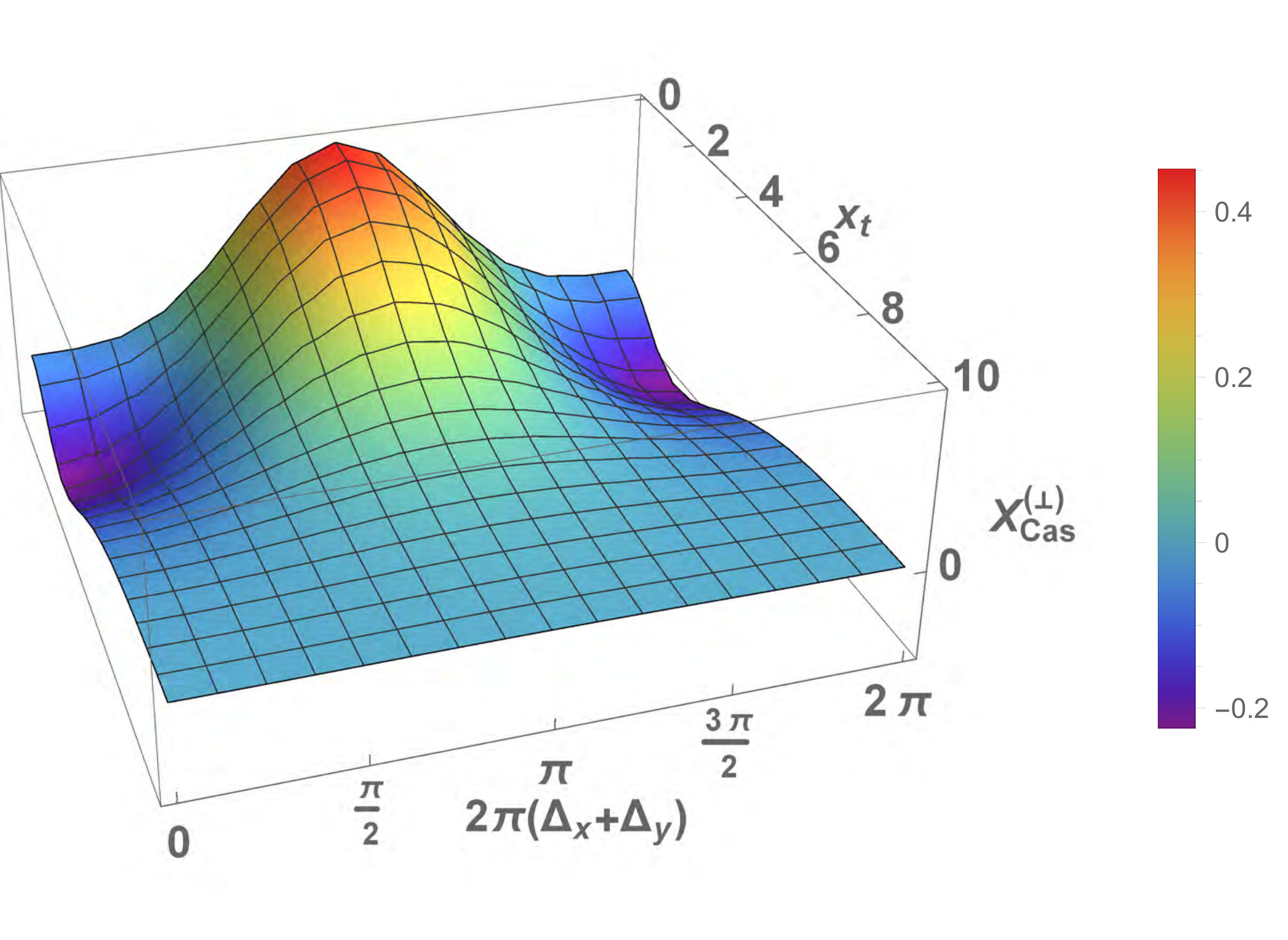}
\caption{(Color online) The scaling function 
$X^{(\perp)}_{\rm Cas}(x_t,x_k=0,x_1,x_L)$
 as a function of $x_t\in (0,10]$ and $\Delta_x+\Delta_y\in [0,1]$ for $x_1=x_L=1$. As wee see, $X^{(\perp)}_{\rm Cas}$ can be both positive and negative, depending on the values of its arguments.}
\label{Fig:3D_G_h1_eq_hL_Legend_total_force_MN}
\end{figure}

The behavior of the scaling function of the total orthogonal Casimir force $X^{(\perp)}_{\rm Cas}(x_t,x_k,x_1,x_L)$ is depicted in Figs. \ref{Fig:3D_G_h1_eq_hL_Legend_total_force} - \ref{Fig:3D_G_h1_not_eq_hL_Legend_total_force_2} for the case when {\it i)} $p\ne  M$ or $q\ne N$ and in the Figs. \ref{Fig:3D_G_h1_eq_hL_Legend_total_force_MN} for the case {\it ii)} $p=M$ and $q=N$ with $x_k=0$. Let us note that in the case {\it i)} the function $X^{(\perp)}_{\rm Cas}$ is symmetric about $x_1$ and $x_L$, while in the case {\it ii)} that is not so. The last implies that when $x_1\ne x_L$ in the case {\it ii)} we have to consider separately the sub-case $x_1\gg x_L$ and $x_1 \ll x_L$. 

\begin{figure}[h]
\includegraphics[width=\columnwidth]{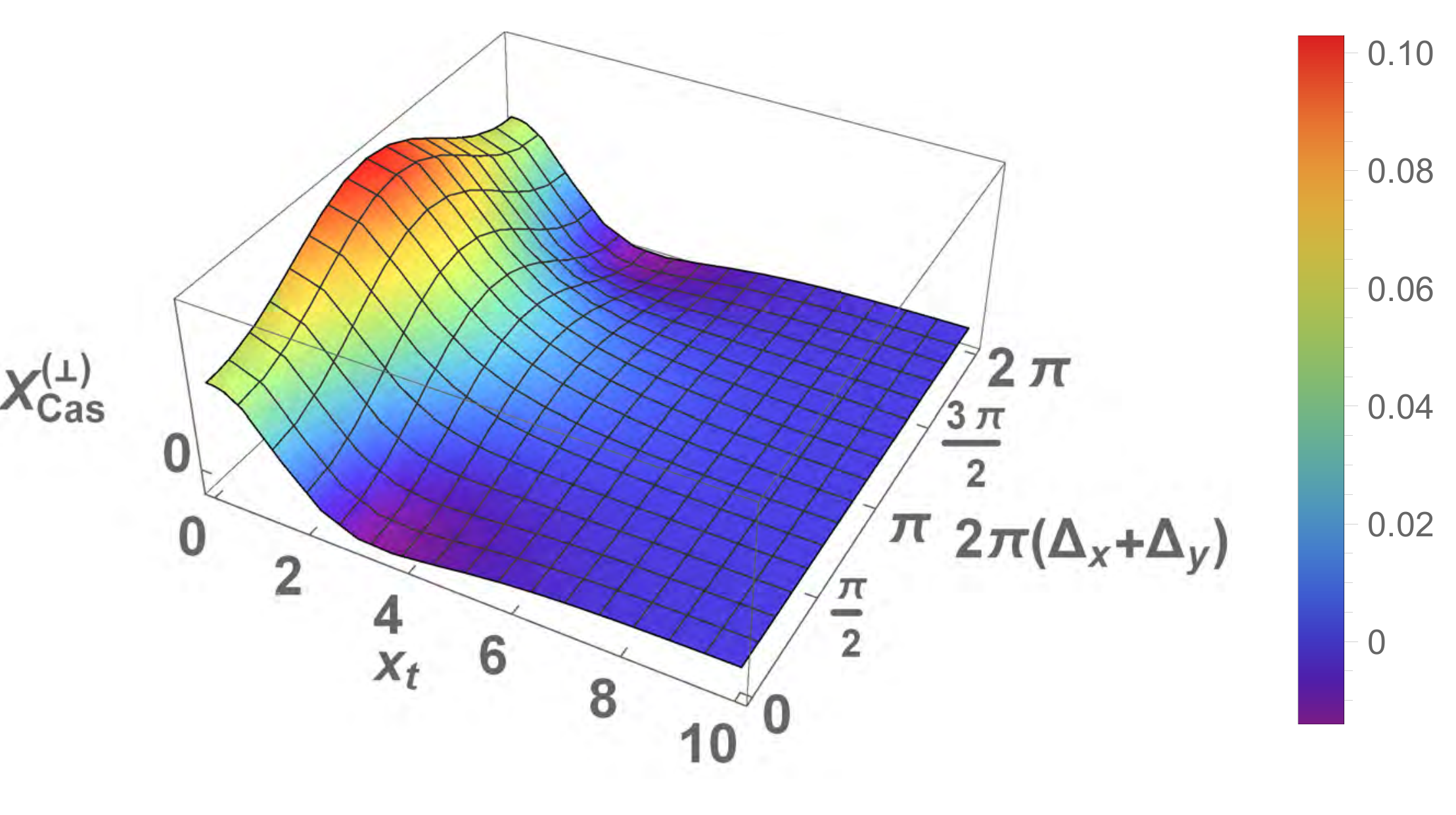}
\caption{(Color online) The scaling function 
$X^{(\perp)}_{\rm Cas}(x_t,x_k=0,x_1,x_L)$
 as a function of $x_t\in (0,10]$ and $\Delta_x+\Delta_y\in [0,1]$  for $x_1=10 x_L=1$. As wee see, the scaling function in that case is predominantly positive.}
\label{Fig:3D_G_h1_not_eq_hL_Legend_total_force_1_MN}
\end{figure}

\begin{figure}[h]
\includegraphics[width=\columnwidth]{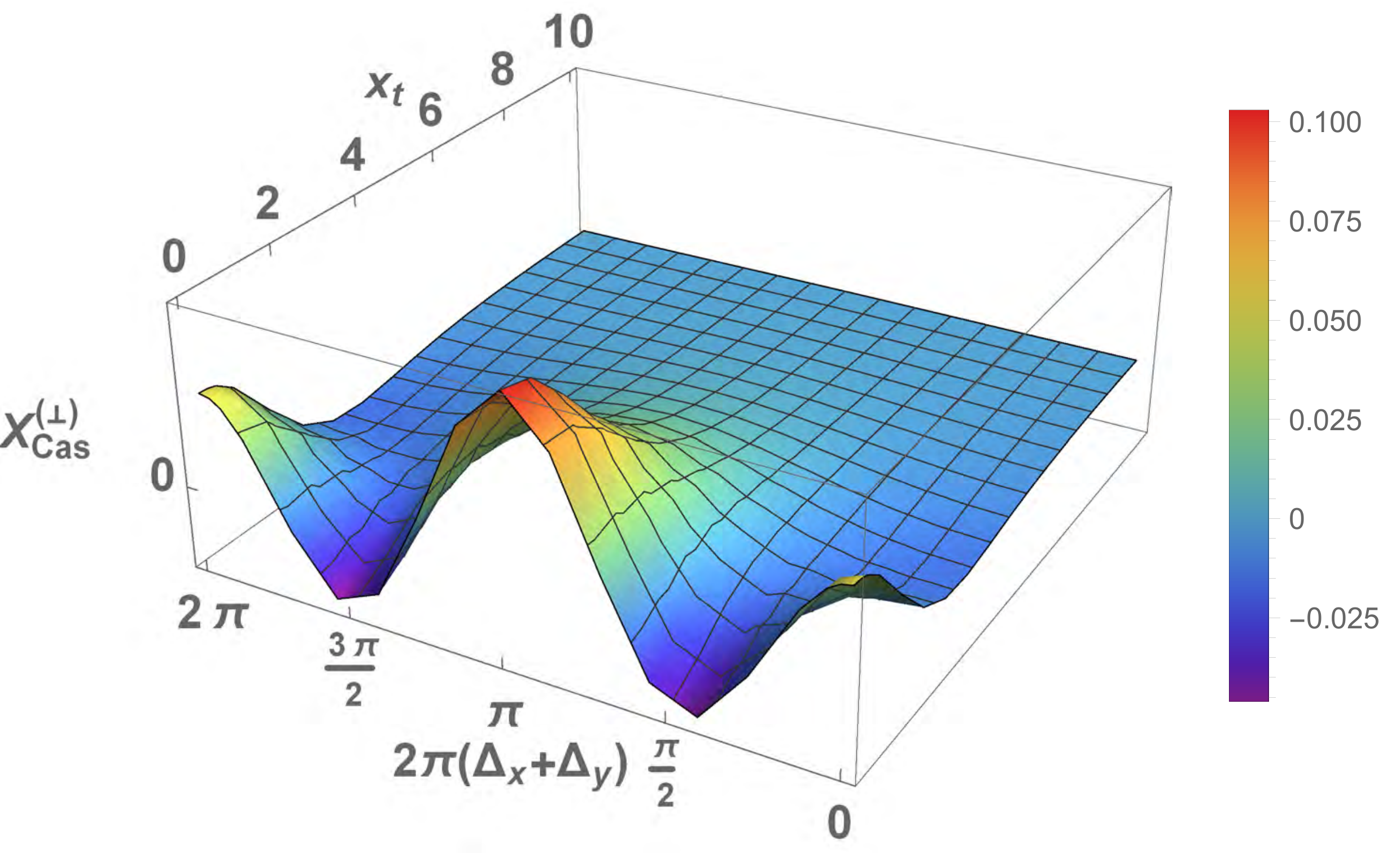}
\caption{(Color online) The scaling function 
$X^{(\perp)}_{\rm Cas}(x_t,x_k=0,x_1,x_L)$
 as a function of $x_t\in (0,10]$ and $\Delta_x+\Delta_y\in [0,1]$  for $10 x_1= x_L=1$. As wee see, the scaling function in that case can be both positive and negative.}
\label{Fig:3D_G_h1_not_eq_hL_Legend_total_force_11_MN}
\end{figure}

\begin{figure}[h]
\includegraphics[width=\columnwidth]{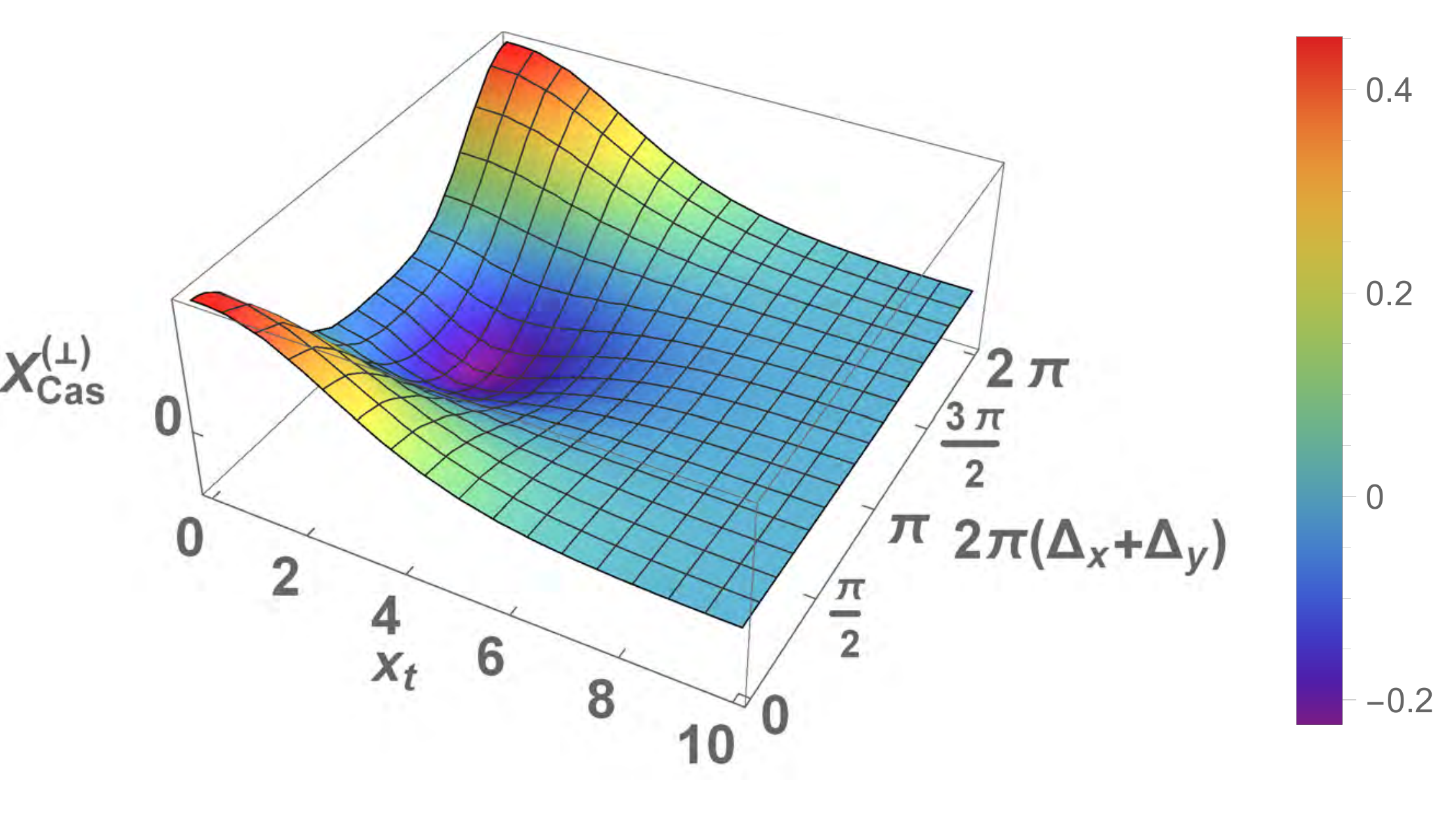}
\caption{(Color online) The scaling function 
$X^{(\perp)}_{\rm Cas}(x_t,x_k=0,x_1,x_L)$
 as a function of $x_t\in (0,10]$ and $\Delta_x+\Delta_y\in [0,1]$  for $x_1= -x_L=1$ or $x_1=-x_L=-1$. As wee see, the scaling function in that case can be both positive and negative.}
\label{Fig:3D_G_h1_not_eq_hL_Legend_total_force_2_MN}
\end{figure}

Figs. \ref{Fig:3D_G_h1_eq_hL_Legend_total_force}  and \ref{Fig:3D_G_h1_eq_hL_Legend_total_force_MN} show the behavior of the force for for equal values of the field  scaling variables $x_1=x_L$. When they are not equal this behavior is visualized in Figs.  \ref{Fig:3D_G_h1_not_eq_hL_Legend_total_force_1} and \ref{Fig:3D_G_h1_not_eq_hL_Legend_total_force_2} for the case {\it i)} and in Figs. \ref{Fig:3D_G_h1_not_eq_hL_Legend_total_force_1_MN}, \ref{Fig:3D_G_h1_not_eq_hL_Legend_total_force_11_MN} and \ref{Fig:3D_G_h1_not_eq_hL_Legend_total_force_2_MN} for the case {\it ii)}. Figs. \ref{Fig:3D_G_h1_not_eq_hL_Legend_total_force_1} and \ref{Fig:3D_G_h1_not_eq_hL_Legend_total_force_1_MN} represent the situation when $x_1\gg x_L$, namely $x_1=10 x_L$, while Figs. \ref{Fig:3D_G_h1_not_eq_hL_Legend_total_force_2} and  \ref{Fig:3D_G_h1_not_eq_hL_Legend_total_force_2_MN} represent the results for the case when $x_1=-x_L=1$. 

The comparison of these figures with Figs. (\ref{Fig:3D_G_h1_eq_hL_Legend}) and (\ref{Fig:3D_G_h1_eq_hL_Legend_MN}) shows, as it might be expected from the data presented in Fig. (\ref{Fig:3d_G_X_Cas_Zero_Field}), that the contribution of $X^{(0,\perp)}_{\rm Cas}(x_t)$ to the overall behavior of the force is quite small, at least in the depicted cases. 

Let us now consider the behavior of the longitudinal Casimir force. 
We first note that it does not have a contribution that is field-independent. Thus, the scaling fuction, which characterizes this force, is given by \eq{eq:Casimir_longit_pq_scaling} and 
\eq{eq:F_long_scaling}. 
\begin{figure}[h]
\includegraphics[width=\columnwidth]{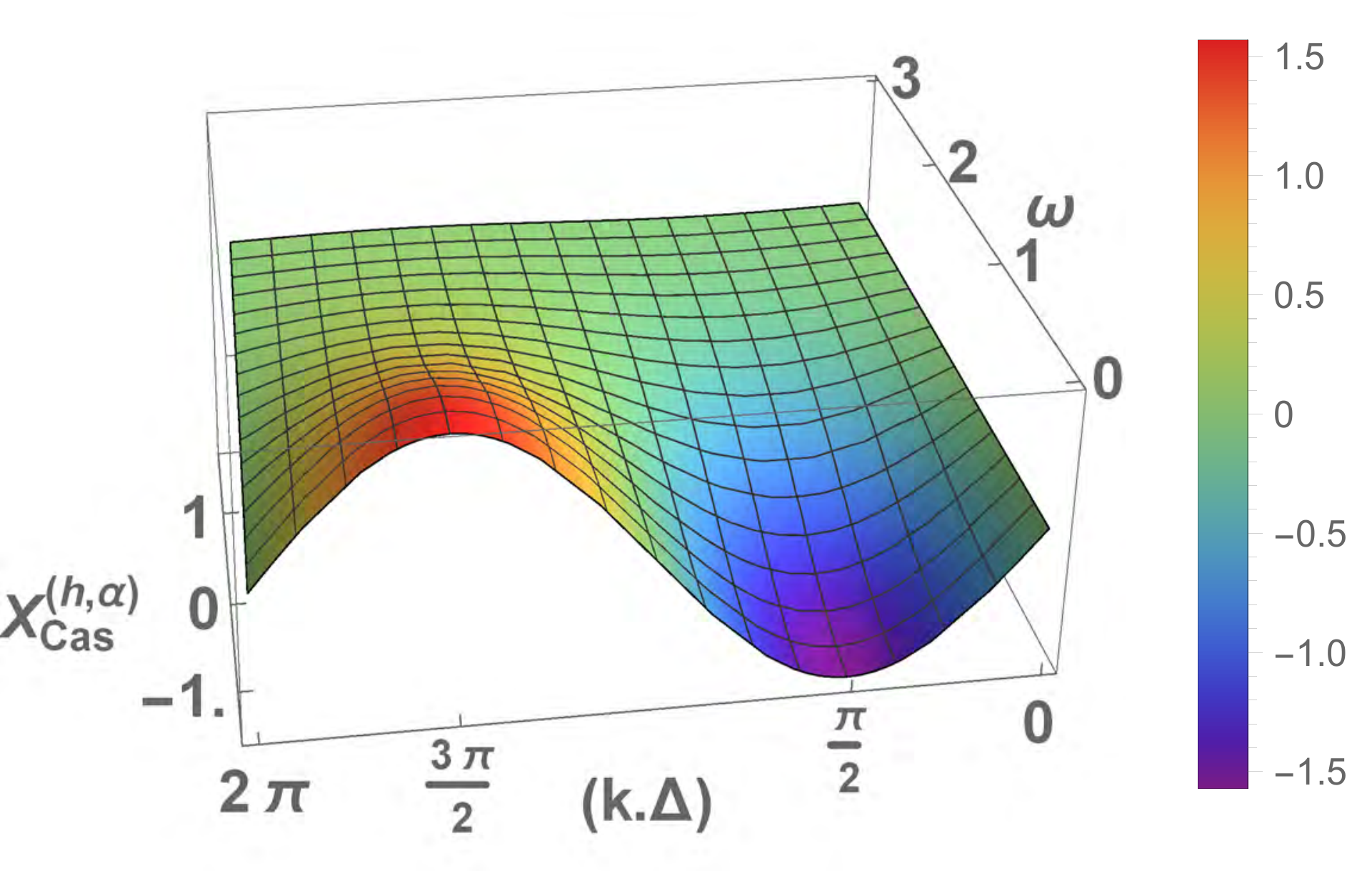}
\caption{(Color online) The scaling function $X^{(h,\alpha)}_{\rm Cas}(w,x_1,x_L)$, see \eq{eq:Casimir_longit_pq_scaling}, as a function of $w\in (0,3]$ and $\left(\mathbf{k.\Delta}\right)\in [0,2\pi]$ for $x_1=x_L=1$. }
\label{Fig:3D_G_h1_eq_hL_Legend_lat}
\end{figure}
\begin{figure}[h]
\includegraphics[width=\columnwidth]{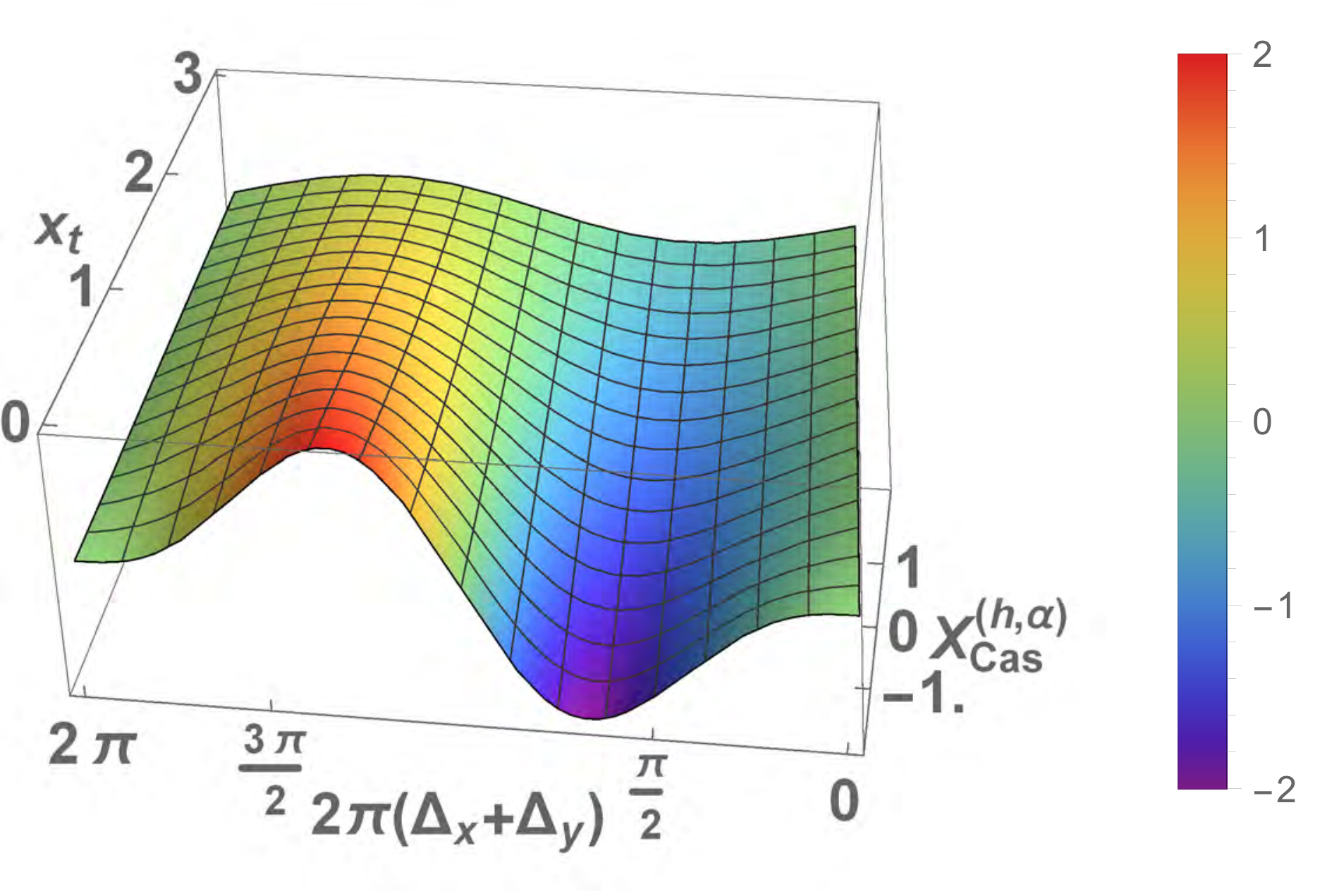}
\caption{(Color online) The scaling function $X^{(h,\alpha)}_{\rm Cas}(x_t,x_1,x_L)$, see \eq{eq:F_long_scaling}, as a function of $w\in(0,3]$ and $\Delta_x+ \Delta_y\in [0,1]$ for $x_1=x_L=1$. Let us remind that in this sub-case $\omega=x_t/2$.  }
\label{Fig:3D_G_h1_eq_hL_Legend_MN_lat}
\end{figure}
Because of the term $\sin (\mathbf{k.\Delta})$, multiplying the expression for the force in the first case, and to $\sin [2 \pi (\Delta_x+ \Delta_y)]$, in the second case, the  scaling function  $X^{(h,\alpha)}_{\rm Cas}$ can be both positive and negative, independently on the values of $x_1$ and/or $x_L$.

\section{The 3d mean-field XY model} \label{sec:3dmf}

\subsection{With infinite surface fields} \label{subsec:infiniteh}

In Ref. \cite{BDR2011} the $XY$ model characterized by the functional
\begin{multline}
{\cal F}\left[ {\bf m};t,L\right]=\int_{-L/2}^{L/2} dz\,\left[\frac{b}{2}\left|\frac{d{\bf m}}{dz}\right|^2+\frac{1}{2}at\left|\textbf{m}\right|^2\right.\\
\left.+\frac{1}{4}g\left|{\bf m}\right|^4\right],
\label{LGenergyfunctional}
\end{multline}
has been studied in the presence of what have been termed twisted boundary conditions. 

\begin{figure}[h]
\includegraphics[width=\columnwidth]{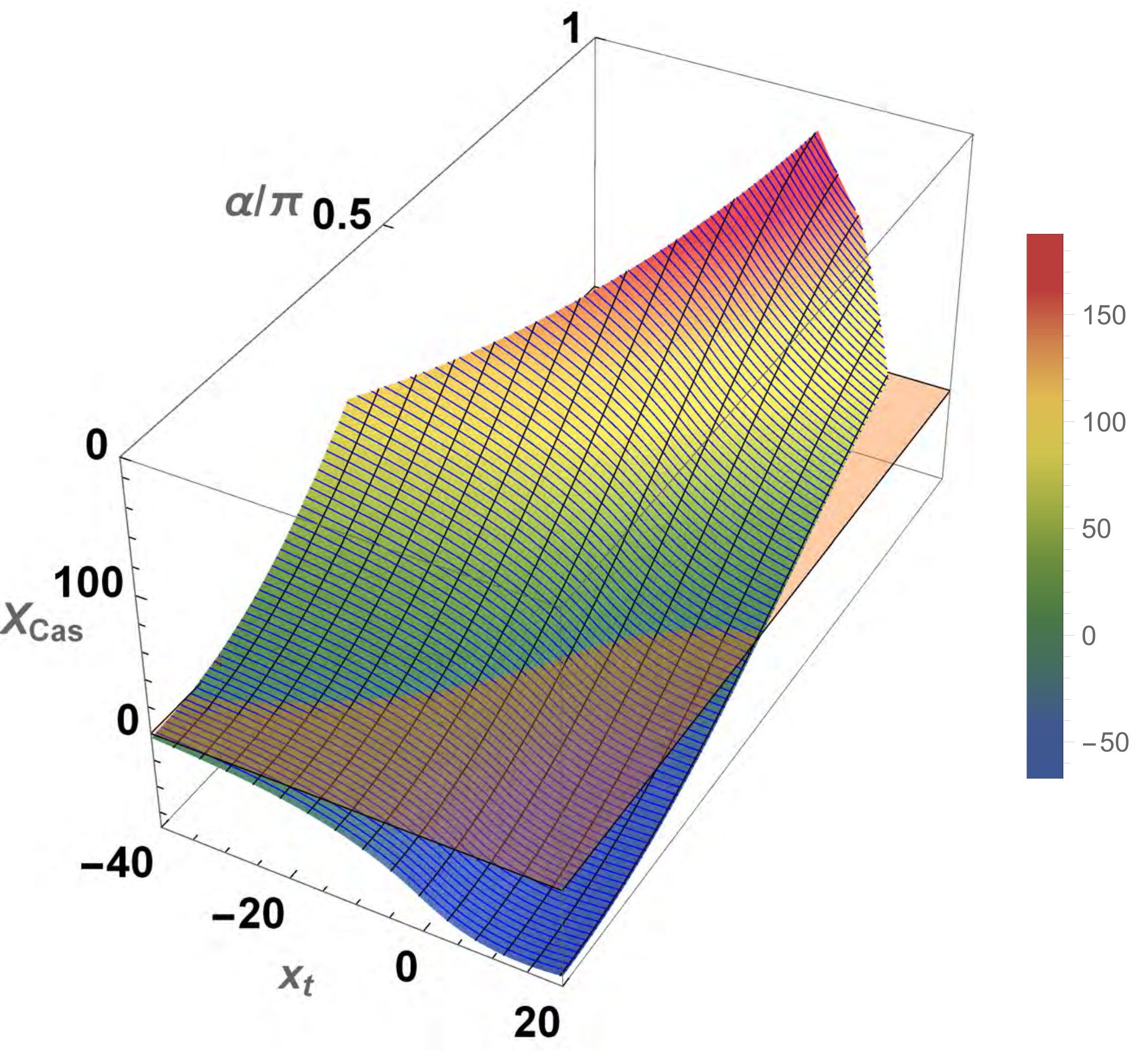}
\caption{(Color online) The scaling function $X^{(\alpha)}_{\rm Cas}(x_t)$ of the $XY$ model under twisted boundary conditions as a function of $x_t$ and $\alpha$ for $h=0$. The plane surface marks the $X^{(\alpha)}_{\rm Cas}(x_t)=0$ value of the force: the force is repulsive above it and attractive below it.}
\label{Fig:3d_Casimir_Force_XY}
\end{figure}
Switching to polar coordinates,
\begin{equation}
{\bf m}(z)=\left(\Phi(z)\cos\varphi(z),\Phi(z)\sin\varphi(z)\right),
\end{equation}
these boundary conditions can convenietly be defined by requiring that \begin{eqnarray}
&\varphi(\pm L/2)=\pm \alpha/2,\nonumber\\
&\Phi(\pm L/2) = \infty,
\label{boundaryconditions}
\end{eqnarray}
i.e., the moments at the boundaries are twisted by an angle $\alpha$ relative to one another. It has been shown that the Casimir force has the form 
\begin{equation}\label{cas}
\beta F_{\rm Cas}(t,L)=\frac{b}{\hat{g}}L^{-4}X_{\rm Cas}^{(\alpha)}(x_t),
\end{equation}
where $\hat{a}=a/b$, $\hat{g}=g/b$, $x_t=\hat{a}t L^{2}$ and 
\begin{equation}\label{casscalingfunctionpandtau}
X_{\rm Cas}^{(\alpha)}(x_t)=\left\{ \begin{array}{cc}
                             X_0^4[p^2- \left(1+\tau \right)],  & x_t \ge 0 \\
                              X_0^4[p^2- \left(1+\tau/2 \right)^2], & x_t \le 0
                            \end{array} \right. .
\end{equation}
Here
\begin{equation}\label{newvarscaling}
\tau=x_t/X_0^2, \, X_0=\int_1^\infty \frac{dx}{\sqrt{(x-1)[x^2+x(1+\tau)+p^2]}}
\end{equation}
\begin{equation}\label{xoonpt}
X_0=\int_1^\infty \frac{dx}{\sqrt{(x-1)[x^2+x(1+\tau)+p^2]}},
\end{equation}
and $p$ is to be determined for any fixed value of $x_t$ so that the twisted spins at the boundary make the prescribed angle $\alpha$.  Let
\begin{equation}\label{roots}
x_\pm=\frac{1}{2} \left[-(\tau +1)\pm\sqrt{(\tau +1)^2-4 p^2}\right]
\end{equation}
be the roots of the quadratic term in the square brackets in the denominator of (\ref{xoonpt}). There are two subcases: {it A)}
the roots are real, and {\it B)} the roots are complex conjugates of each
other.

{\it A)} The roots $x_\pm$ are real. Then
\begin{equation}\label{x0det}
X_0=\frac{2}{\sqrt{1-x_-}}K\left[\sqrt{\frac{x_+-x_-}{1-x_-}} \right]
\end{equation}
and 
\begin{multline}\label{alphaasafunctionoftherootsfinal}
\alpha=\frac{\sqrt{|x_- x_+|} X_0}{x_-} \bigg\{1\\
-\frac{2}{X_0 \sqrt{1-x_-}} \Pi \left[\frac{x_-}{x_- -1},\sqrt{\frac{x_+-x_-}{1-x_-}} \right]\bigg\}.
\end{multline}
We note that 
\begin{equation}\label{relxtp}
\tau=-1-x_--x_+,\qquad p=\sqrt{|x_- x_+|}.
\end{equation}

{\it B)} The roots $x_\pm$ are complex.

One has 
\begin{equation}\label{cx0det}
X_0=\frac{2}{\sqrt{r}} K\left(w\right),
\end{equation}
and 
\begin{multline}\label{calphaasafunctionoftherootsfinal}
\alpha = \frac{p X_0}{1-r}+\frac{4p}{r^2-1} \sqrt{\frac{r}{1-w^2}}\\
\times\Pi\left[\left(\frac{r-1}{r+1}\right)^2,\frac{w}{\sqrt{w^2-1}}\right].
\end{multline}
where 
\begin{eqnarray}\label{notations1}
r\equiv r(x_-,x_+)&=&\sqrt{(1-x_-)(1-x_+)} \nonumber\\
&=& \sqrt{2+\tau+p^2},
\end{eqnarray}
and
\begin{multline}\label{notations2}
w^2\equiv w^2(x_-,x_+) = \frac{1}{2}+\frac{\frac{x_- + x_+}{2}-1}{2
   \sqrt{(1-x_-) (1-x_+)}}\\
   = \frac{1}{2}\left(1-\frac{3+\tau}{2\sqrt{2+\tau+p^2}} \right).
\end{multline}
The scaling function $X^{(\alpha)}_{\rm Cas}(x_t)$ of the $XY$ model under twisted boundary conditions as a function of $x_t$ and $\alpha$ is shown in Fig. \ref{Fig:3d_Casimir_Force_XY}.
We recall that, as shown in Ref. \cite{BDR2011} the asymptotic expression for $X_{\rm Cas}^{(\alpha)}(x_t)$
\begin{equation}\label{casscalingfunctionasympXY}
X_{\rm Cas}^{(\alpha)}(x_t)\simeq \frac{1}{2}  \alpha^2 \left[|x_t|+4
 \sqrt{2|x_t|}+\frac{1}{2} \left(48-3 \alpha ^2\right)\right],
\end{equation}
when $x_t\to-\infty$. According to \eq{cas} the last implies that in this regime 
\begin{equation}
\label{casMFas}
\beta F_{\rm Cas}(t,L)\simeq \frac{1}{2} \alpha^2 \frac{b}{\hat{g}}   |x_t|L^{-4} = \frac{1}{2}\frac{a b}{g} \alpha^2  |t| L^{-2},
\end{equation}
i.e., its leading behavior is of the order of $L^{-2}$ there due to the existence of helicity within the system. 

\subsection{With finite surface fields}

The model described immediately above constrains the spins at the surface of the film to point in particular directions. The physical realization of a such a system is much more likely to be one in which the spins at the surfaces to be under the influence of finite surface fields. Here, we consider a model for such a system. In order to do so, we employ the approach utilized in Section II of  \cite{BDR2011}, in which the spin system occupies sites on a lattice that is infinite in extent in two directions and that consists a finite number of layers (here labeled 1 to $L$) in the third dimension. We impose  surface fields that couple in the standard way to the spins on the leftmost layer, labeled 1, and the rightmost layer, labeled $L$. The magnitude of each of those fields is $h_s$, and the angle between them is $\alpha$. In our mean field approach, the free energy is minimized by adjusting the expectation value of the amplitude and direction of the spins in each layer. The Casimir force follows from the difference between the free energies with $L$ and $L+1$ layers; because of the numerical nature of the free energy results, we are unable to take the derivative with respect to film thickness, as in Section \ref{sec:continuum}.

We find that the Casimir force is consistent with the following scaling form
\begin{equation}
F_{\rm Cas} = L^{-4}f(tL^2, h_cL) \label{eq:3dmf1}
\end{equation} 
where $t$ is the bulk reduced temperature.  Furthermore, for small enough $h_c$ and $t$ higher than the value at which the film orders spontaneously, the function $f$ on the right hand side of (\ref{eq:3dmf1}) has the form
\begin{equation}
f(tL^2, h_cL) = f_0(tL^2) + f_1(tL^2) \left(h_cL \right)^2  + O\left( \left(h_cL \right)^4\right)\label{eq:3dmf2}
\end{equation}
Because of this, it is possible to envision for small $h_s$ the behavior of the Casimir force that one encounters in the Gaussian model. 

\begin{figure}[htbp]
\begin{center}
\includegraphics[width=3in]{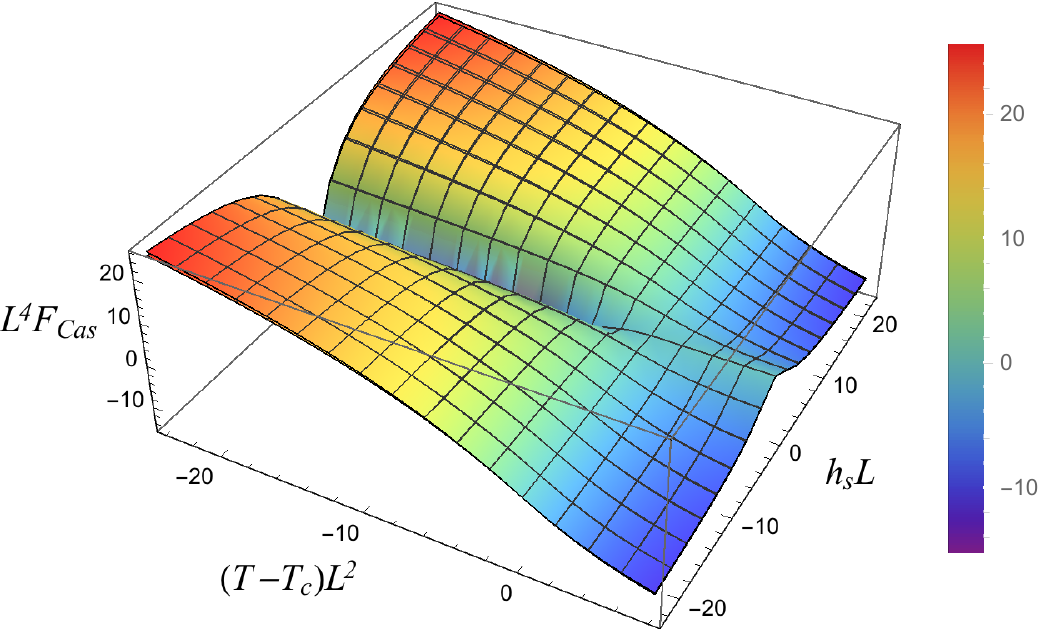}
\caption{(Color online) Scaled Casimir force, $L^4F_{\rm Cas}$, as a function of the scaled reduced temperature, $tL^2$ and scaled surface field amplitude, $h_sL$. The number of layers in the two films are $L=50$ and $L=100$, and $\alpha$, the angle between the surface fields, is $\pi/3$. The difference between the two plots is barely discernible, indicating that the difference between the scaling function for $L=50$ and the infinite $L$ limit is quite small. }
\label{fig:3dmfplot1}
\end{center}
\end{figure}

Figure \ref{fig:3dmfplot1} is a plot of the scaled Casimir force versus the scaled reduced temperature and scaled surface fields for two values of the film thickness, $L$. The perspective highlights the departure from the behavior in (\ref{eq:3dmf2}) that occurs when the temperature is sufficiently far below the bulk critical temperature that the moments in the film order spontaneously. The films in question consists of $L=50$ and $L=100$ layers, and the angle between the two surface fields is $\alpha = \pi/3$. As is clear from the figure, the difference between the two plots is quite small. 

\begin{figure}[htbp]
\begin{center}
\includegraphics[width=3in]{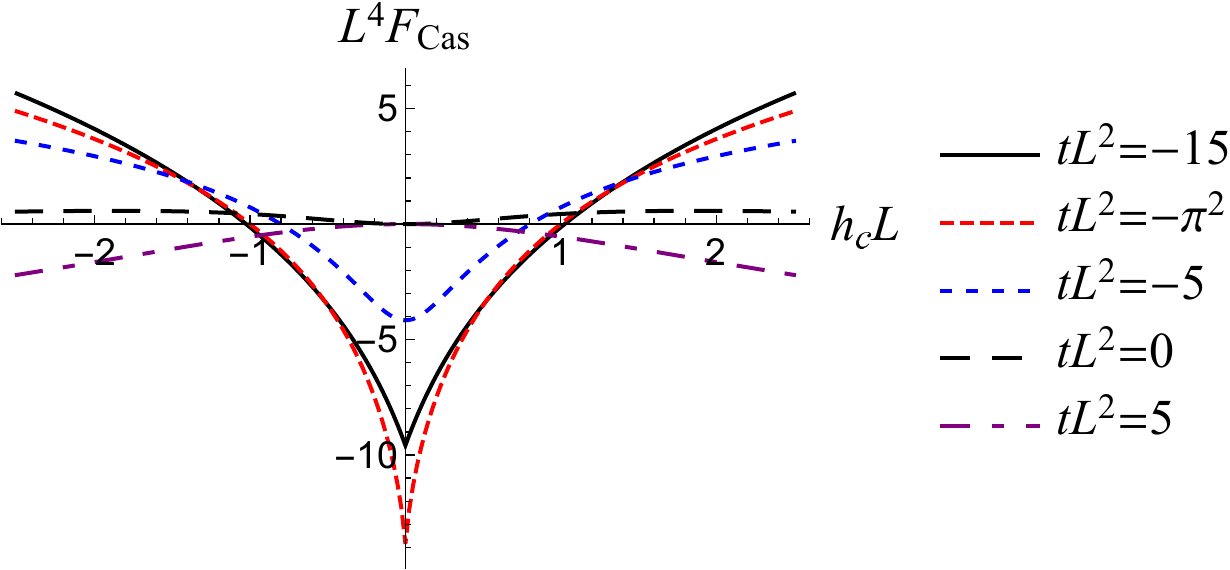}
\caption{(Color online) Scaled Casimir force, $L^4 F_{\rm Cas}$, as a function of the scaled surface field, $h_sL$ for various scaled reduced temperatures, $tL^2$. Here, $L=50$ and $\alpha= \pi/3$. When $tL^2>-\pi^2$ the small $h_s$ dependence of the Casimir force is quadratic, consistent with (\ref{eq:3dmf2}). Below that value of the scaled reduced temperature, the small $h_s$ dependence is linear in the absolute value of that quantity, as exemplified by the curve for $tL^2=-15$. }
\label{fig:3dmfplot2}
\end{center}
\end{figure}

\begin{figure}[htbp]
\begin{center}
\includegraphics[width=3in]{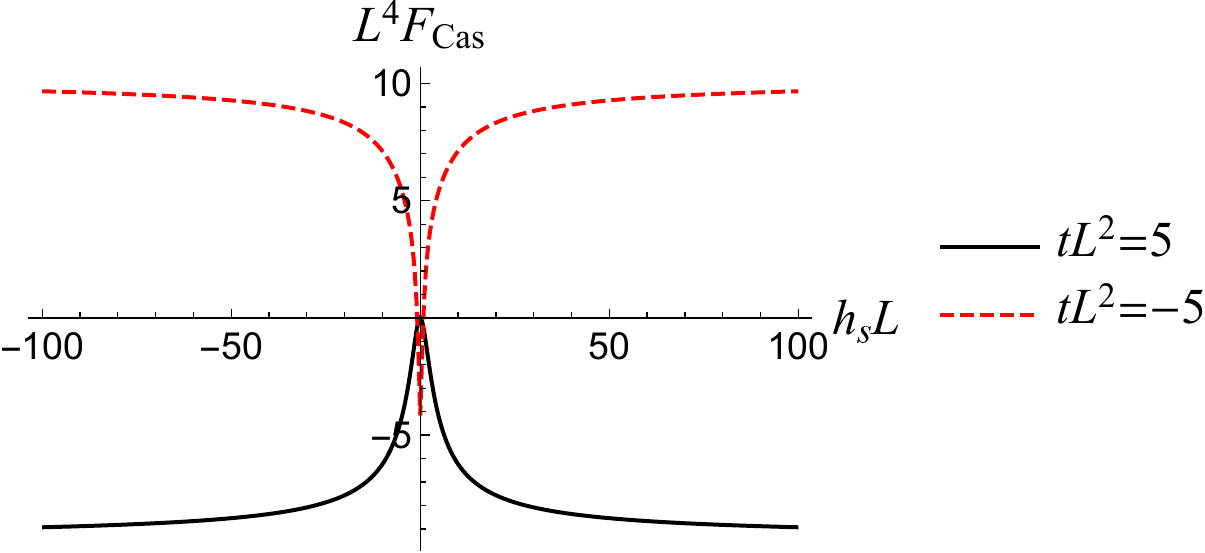}
\caption{(Color online) Dependence of the scaled Casimir force on the scaled surface field for two values of scaled reduced temperature above the point, $tL^2=-\pi^2$, at which spontaneous ordering occurs in the film. Here, $L=50$ and $\alpha = \pi/3$. The plots illustrate the saturation of the influence of the surface fields, at odds with the amplification effect seen in Section  \ref{sec:Gaussian}. The figure  also illustrates the fact that the Casimir force can change change sign as the temperature is varied. This is due to the fact that there is a range of temperatures  below the bulk critical temperature in which the bulk system orders while the film remains disordered. For $T>T_c$ both the bulk and the finite system are disordered. For $|h_s|\gg 1$ the Casimir force approaches its value for fixed boundary conditions, the case considered in Subsection \ref{subsec:infiniteh}.}
\label{fig:3dmfplot3}
\end{center}
\end{figure}

\begin{figure}[htbp]
	\begin{center}
		\includegraphics[width=3in]{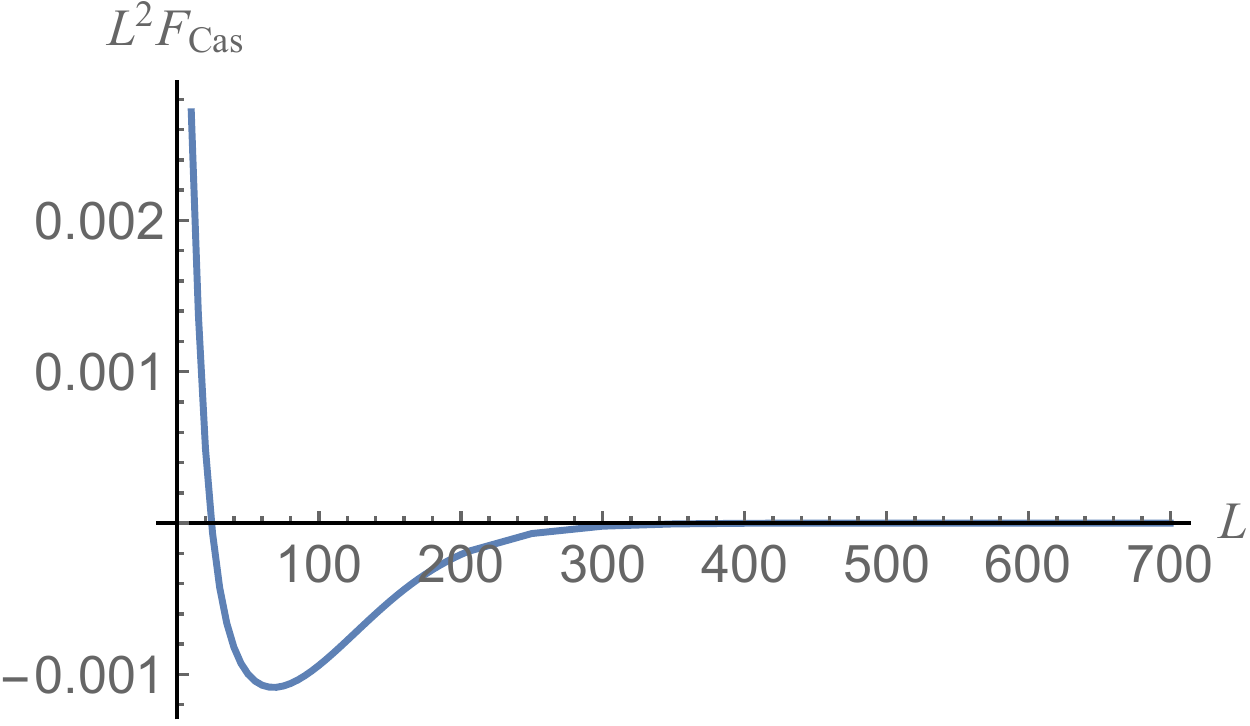}
		\caption{(Color online) Scaled Casimir force, $L^2 F_{\rm Cas}$, as a function of $L$,  for fixed values of temperature $t=0.001$, helicity $\alpha= \pi/3$ and value of the surface field amplitude $h_s=0.1$.  }
		\label{fig:above_tc}
	\end{center}
\end{figure}

As indicated in Fig. \ref{fig:3dmfplot1}, $L=50$ is sufficiently large that the difference between the function and the scaling limit is quite small. Figure \ref{fig:3dmfplot2} illustrates the dependence of the scaled Casimir force on the scaled surface field amplitude for various values of the scaled reduced temperature. 
For all reduced temperatures greater than $-\pi^2$, the initial dependence on scaled surface fields is quadratic, consistent with (\ref{eq:3dmf2}). In fact for temperatures at and above the bulk critical temperature ($t \ge 0$) the second term in the right hand side of (\ref{eq:3dmf2}) is the leading non-zero contribution to that expansion. This is consistent with the amplification of the Casimir force that one finds in the Gaussian model---see Section \ref{sec:Gaussian}. However, such amplification only occurs when there is spontaneous ordering in the film. Figure \ref{fig:3dmfplot3} shows the scaled Casimir force as a function of the scaled surface field for $tL^2=5$ and $tL^2=-5$, above and below the bulk transition but above the threshold for film ordering. This plot illustrates the saturation of the Casimir force when the reduced temperature is above the threshold for film ordering, $tL^2=-\pi^2$.  
The Casimir force changes sign as $L$ increases for fixed $\alpha$, $T$ and $h_s$. This is displayed in Fig. \ref{fig:above_tc}. 
We also note that the force changes sign for moderate values of $L$. It can readily be established that the overall behavior of the Casimir force is 
in accord with Eq. (\ref{eq:3dmf2}); see, for instance, Fig. \ref{fig:3dmfplot3}.

\begin{figure}[htbp]
\begin{center}
\includegraphics[width=3in]{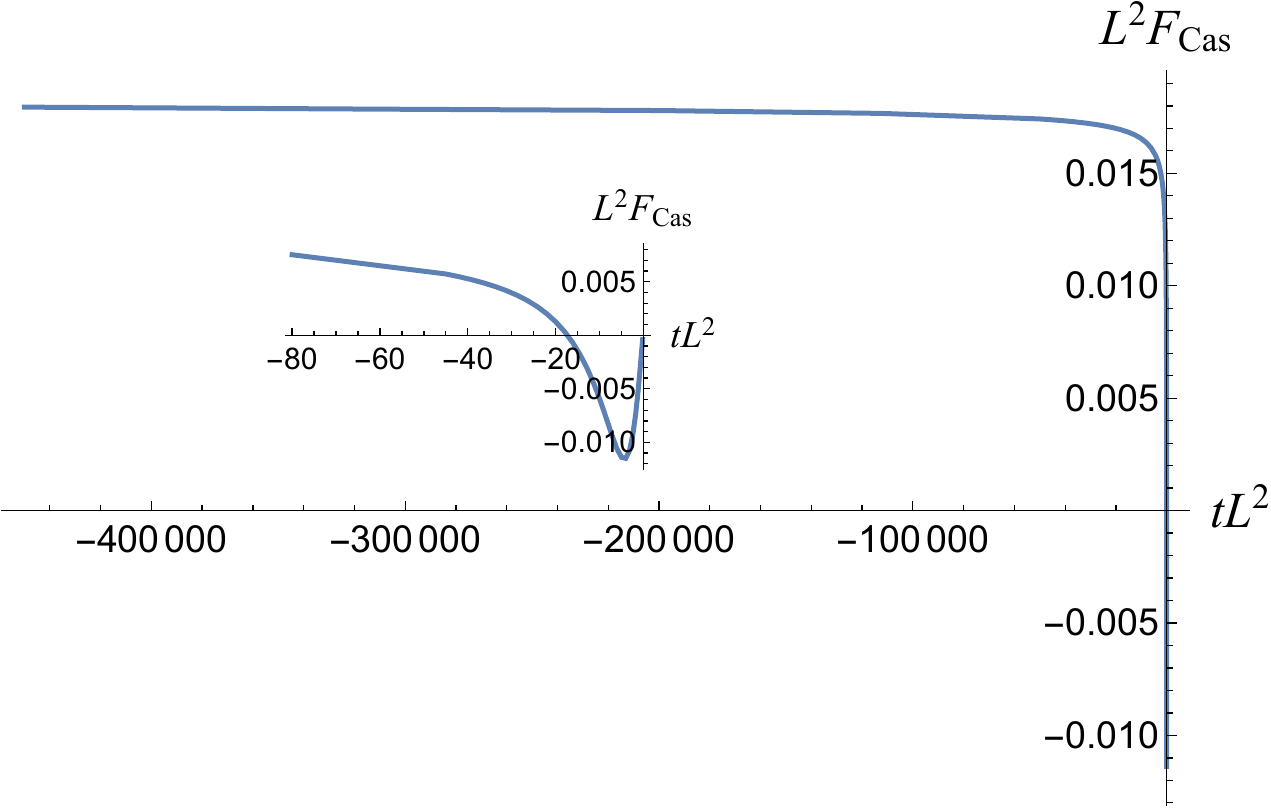}
\caption{(Color online) Illustrating the $L$ dependence of the Casimir force for a negative value of reduced temperature, $t=-0.05$ with surface field amplitude $h_s=0.05$ and $\alpha=\pi/3$. The plot is generated by varying the film thickness $L$ for fixed values of $t$, $h_s$ and $\alpha$. The large graph shows how $L^2F_{\rm Cas}$ varies over an extended range of film thicknesses $L$, and the inset shows the $L$ dependence over a much smaller range.}
\label{fig:3dmfplot4}
\end{center}
\end{figure}

If spontaneous ordering is possible, then amplification of the Casimir force does occur. Figure \ref{fig:3dmfplot4} plots the newly scaled Casimir force $L^2F_{\rm Cas}$ against system size $L$, illustrating the enhanced force amplitude as a function of system size, $L$, expressed in terms of the scaled variable $tL^2$. Here, the reduced temperature is fixed at $t=-0.05$, while the surface field amplitudes are set to $0.05$, $\alpha = \pi/3$ and the system size varies from $L=2$ to $L=3,000$. The behavior displayed is a direct result of the energy stored in the helical spin configuration, a response to the surface fields that are tilted with respect to each other. 
Of additional interest in this plot is the variation of the Casimir force for smaller values of $L$, shown in the inset. Note  the change in the sign of the Casimir force. A Casimir force going as $L^{-2}$ is consistent with the energy associated with a helicity modulus, which is natural given that the $XY$ system supports such a modulus in the regime in which it spontaneously orders. In this case the surface fields play the essential role of enforcing a helical structure on the order parameter when spontaneous ordering occurs. 

\begin{figure}[htbp]
\begin{center}
\includegraphics[width=3in]{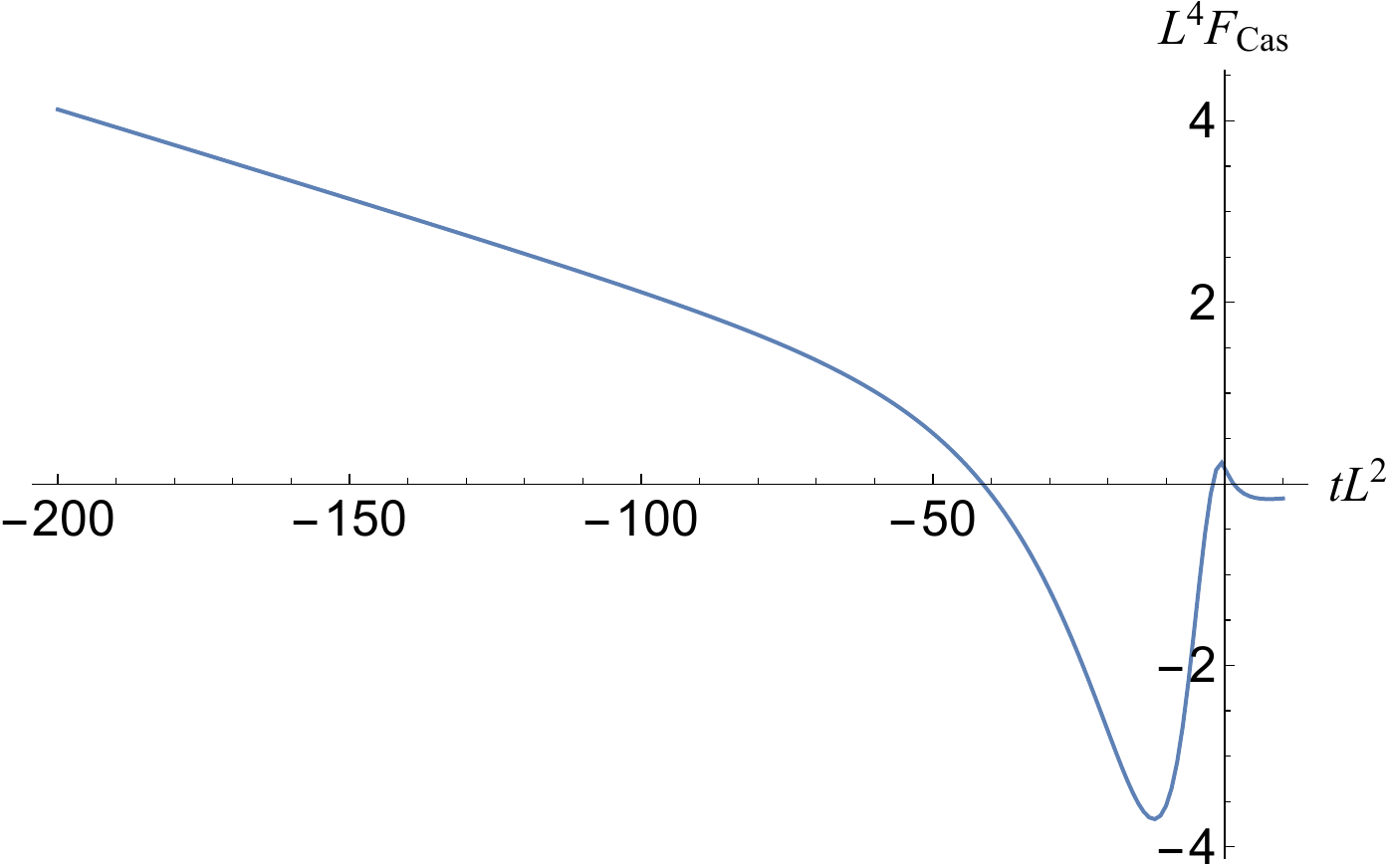}
\caption{(Color online) The scaled Casimir force, $L^4 F_{\rm Cas}$, as a function of the scaled variable $tL^2$. The thickness of the film is $L=50$, the surface field amplitudes have been set to 0.01 and the angle between them, $\alpha$, is $\pi/3$. }
\label{fig:3dmfplot5}
\end{center}
\end{figure}

The enhanced Casimir force is consistent with the scaling form of (\ref{eq:3dmf1}). Figure \ref{fig:3dmfplot5} displays the dependence of the scaled Casimir force $L^4 F_{\rm Cas}$ on the scaled variable $tL^2$.
An important feature of this plot is its linear dependence on the scaled reduced temperature when it is sizable and negative. This leads to an overall  $L$ dependence going as $L^{-2}$. Another significant property of the critical Casimir force plotted in Fig. \ref{fig:3dmfplot5} is its change in sign in the vicinity of the bulk critical point. In this sense, the Casimir force is tunable---and can be changed from attractive to repulsive---through a variation in temperature.

\begin{widetext}

\begin{figure}[h!]
\begin{center}
\includegraphics[width=7in]{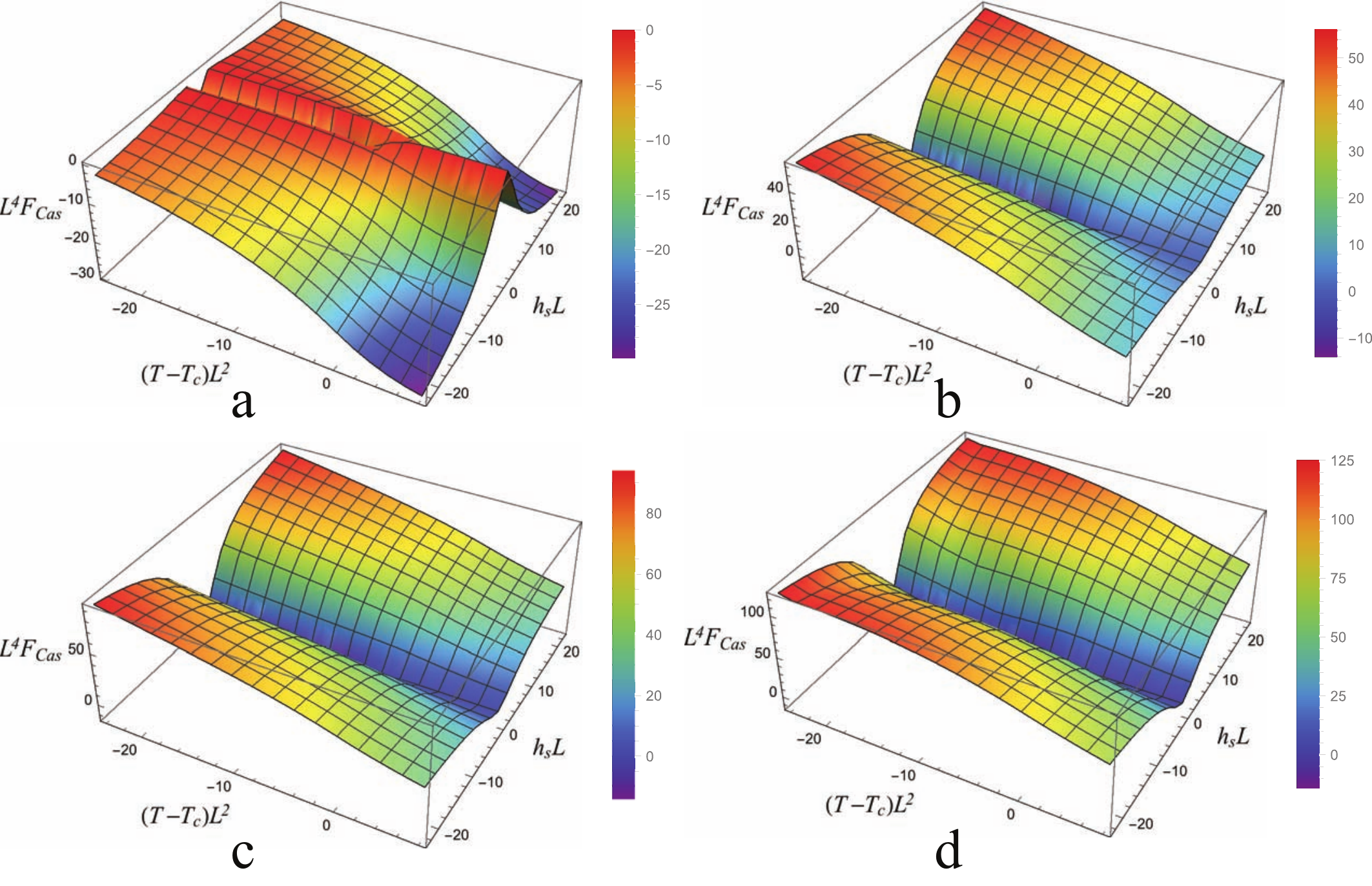}
\caption{(Color online) Scaled Casimir force, $L^4F_{\rm Cas}$, as a function of the scaled reduced temperature, $tL^2$ and scaled surface field amplitude, $h_sL$. The number of layers in the film is $L=50$. The values of $\alpha$, the angle between the surface fields are, reading left to right and then top to bottom are {\bf a}: 0, { \bf b}: $\pi/2$, { \bf c}: $2 \pi/3$ and { \bf d}: $\pi$. }
\label{fig:3dmfplot6}
\end{center}
\end{figure}
\end{widetext}

Finally, Fig. \ref{fig:3dmfplot6} displays the dependence of the scaled Casimir force, $L^4F_{\rm Cas}$, on scaled reduced temperature, $tL^2$ and scaled surface field amplitude, $h_sL$ for a variety of values of the angular difference, $\alpha$, between the two surface fields. As shown in the plots, when $\alpha$ increases from $0$ to $\pi$ the minimum of the force becomes shallower and the region of parameters $tL^2$ and $h_sL$ in which the force is repulsive expands. We also note that the amplitude of the force  for any fixed combination of the parameters $tL^2$ and $h_sL$ is a monotonically increasing function of $\alpha$. The force is attractive in the whole region of $h_sL$ and $tL^2$ values only for $\alpha=0$.

 \pagebreak

\section{Discussion and concluding remarks} \label{sec:conclusions}
The Casimir force has provided an unexpectedly rich and varied set of phenomena for study and potential exploitation. In this paper, we have attempted to demonstrate that interactions between the bounding system and the media that supports the Casimir force allow for the possibility of utilizing those interactions, here parameterized as surface fields, to control---and in certain cases greatly amplify---that force. Our focus has been the critical Casimir force, but a number of our results extend far beyond the critical regime. We find that the angle between surface fields can significantly affect the magnitude and the sign of the Casimir force, that variations in temperature can also have such an effect, and that the strength of the critical Casimir force can undergo substantial amplification as a consequence of the application of surface fields. Such fields represent a useful and likely accurate quantification of the action of modifications of the structure or composition of bounding surfaces in the medium giving rise to the Casimir force. Thus, the results presented here could well be utilized or expanded upon to motivate experimental investigations of the effects of surface patterning on the Casimir force. 

The key findings reported here are twofold. First, the combination of helicity and surface fields allows for the manipulation of both the sign and the amplitude of the Casimir force. In certain circumstances---particularly when the system supports helicity in the bulk---the force can be greatly amplified in magnitude. The second finding is that the expressions describing the Casimir force are consistent with the expectations of finite size scaling, as embodied in Eqs. (\ref{eq:F_Cas_scaling}), (\ref{eq:F_Cas_scaling_Heis}), (\ref{eq:F_Cas_no_field}), (\ref{eq:sc_funct_field}), (\ref{eq:sc_funct_field_long}), (\ref{cas}) and (\ref{eq:3dmf1}).

One possible setting for an experimental study might be a nematic liquid crystal film. Here, the order parameter is quadrupolar, rather than dipolar as in the case of the $XY$ or Heisenberg models, but the continuous symmetry with respect to rotation of the order parameter is nevertheless in the same general class as in the systems considered here. In fact, a class of Liquid Crystal Display (LCD) devices operates on the basis of inducing of a helical structure in liquid crystalline films  \cite{LCDGray}. It is also possible that the results reported here are applicable to the case of a liquid Helium film in the superfluid state in which a temperature gradient exists between the substrate on which the film has condensed and that gas phase bordering its free surface. Such a temperature gradient induces flow in the superfluid component, which entails a rotation of the superfluid wave function in the complex plane \cite{[{R. P. Feynman in }] Gorter,Ginz_PIt}.

The models investigated here are unlikely to be directly realized in nature, either because of their low dimensionality,  or because they neglect important phenomena such as saturation of the order parameter as in the Gaussian model or are based on approximations, such as the mean field theory. Nevertheless, we are confident in the the overall import of our results: that surface fields and helicity in the medium that generates the Casimir force are likely to prove quite significant as experimentally accessible modifiers of that force. How those surface fields are to be generated will vary from system to system, but there is every reason to anticipate that ways will be found and that the result will be a greater insight into the Casimir force and, one hopes, new and useful applications of this interaction. 

\acknowledgements{D.D. gratefully acknowledges the  financial support via contract DN02/8 of Bulgarian NSF. J. R. is pleased to acknowledge support from the NSF through DMR Grant No. 1006128}

\appendix
\section{Calculation of the free energy for the 1d XY model with boundary fields} \label{app:XY}

The simplest way we are aware of for calculation of the free energy of the $1$d XY model is based on the identities \cite{AS,NIST2010}
\begin{equation}
\label{eq:Bessel_idemtity}
e^{z\cos \theta}=\sum_{n=-\infty}^{\infty} e^{i n \theta} I_{n}(z)
\end{equation}
and
\begin{equation}
\label{eq:delta_function_idemtity}
\delta_{n,0}=\frac{1}{2\pi}\int_{0}^{2\pi} e^{i n \theta}  d\theta.
\end{equation}
From Eqs. \eqref{eq:system_angles}  and \eqref{eq:free_energy} one then obtains
\begin{eqnarray}
\label{eq:free_energy_calculating}
\lefteqn{\exp \left(-\beta  F_N\right)=\int_{0}^{2\pi}\prod_{i=1}^{N} \frac{d\varphi_i}{2\pi}\times}\\&& \sum_{n_1=-\infty}^{\infty} e^{i n_1 (\psi _1-\varphi _1)} I_{n_1}(h_1)\sum_{n_2=-\infty}^{\infty} e^{i n_2 (\varphi _1-\varphi_2)} I_{n_2}(K)
\nonumber \times\\
&& \cdots \times \sum_{n_{N-1}=-\infty}^{\infty} e^{i n_{N-1} (\varphi _{N-1}-\varphi_N)} I_{n_{N-1}}(K)\times \\&& \sum_{n_N=-\infty}^{\infty} e^{i n_N (\varphi _{N}-\psi_N)} I_{n_{N}}(h_N),
\end{eqnarray}
wherefrom, using \eq{eq:delta_function_idemtity}, one obtains \eq{eq:free_energy_calculated}. Obviously, \eq{eq:free_energy_calculated} can be written in the form 
\begin{eqnarray}
\label{eq:free_energy_cos}
\lefteqn{\exp \left(-\beta  F_N\right)=I_0 \left(h_1\right) I_0(K)^{N-1} I_0\left(h_N\right) \times} \\
&& \left[1+2\sum _{k=1 }^{\infty } \cos \left(k \psi\right) \frac{I_k\left(h_1\right)}{I_0\left(h_1\right)} \left(\frac{I_k(K)}{I_0(K)}\right)^{N-1} \frac{I_k\left(h_N\right)}{I_0\left(h_N\right)}\right].\nonumber
\end{eqnarray}
From \eq{eq:free_energy_cos} for the total pressure 
\begin{equation}
\label{eq:def_Ftot}
\beta  F_{\text{tot}}=-\frac{\partial}{\partial N}[\beta  F_N]
\end{equation}
exerted by the end points on the system one then obtains
\begin{widetext}
\begin{equation}
\label{tot}
\beta  F_{\text{tot}}=\ln I_0(K)+
\frac{ 2\sum _{k=1}^{\infty } 
\cos \left[k (\psi_1-\psi_2)\right] \log \left[\frac{I_k(K)}{I_0(K)}\right]\frac{I_k\left(h_1\right)}{I_0\left(h_1\right)} \left(\frac{I_k(K)}{I_0(K)}\right)^{N-1} \frac{I_k\left(h_N\right)}{I_0\left(h_N\right)}
}{1+2 \sum _{k=1}^{\infty } \cos \left[k (\psi_1-\psi_2)\right] \frac{I_k\left(h_1\right)}{I_0\left(h_1\right)} \left(\frac{I_k(K)}{I_0(K)}\right)^{N-1} \frac{I_k\left(h_N\right)}{I_0\left(h_N\right)}},
\end{equation}
\end{widetext}
wherefrom one immediately derives the expression \eqref{FCas}
 for the Casimir force given in the main text. 
 
Using the Poisson identity 
\begin{equation}
\label{eq:Poisson}
\sum _{k=-\infty }^{\infty } \exp \left(i
   k a-k^2 b\right)=\sqrt{\frac{\pi }{b}}
   \sum _{n=-\infty }^{\infty }
   \exp{\left[-\frac{(2 \pi  n+a)^2}{4 b}\right]}
\end{equation}
one can derive expressions for the scaling function of the Casimir force convenient for values of the scaling variable $x$ ranging from moderate to large values of $x$ and one convenient for small values of $x$. 

\section{Calculation of the free energy for the 1d Heisenberg model with boundary fields}
\label{app:Heisenberg}

Let us write the vectors in \eq{eq:def_1d_Ham} in spherical coordinates supposing the spin chain to be along the $x$ axis. One has
\begin{eqnarray}
\label{def}
{\mathbf H}_1 &=& H_1\left \{\sin\varphi_1^h\cos\theta_1^h,\sin\varphi_1^h\sin\theta_1^h,\cos\varphi_1^h\right \} \\
{\mathbf H}_N &=& H_N\left \{\sin\varphi_N^h\cos\theta_N^h,\sin\varphi_N^h\sin\theta_N^h,\cos \varphi_N^h\right \}
\nonumber \\
{\mathbf S}_i &=& \left \{\sin\varphi_i\cos\theta_i,\sin \varphi_i\sin \theta_i,\cos\varphi_i\right \}, i= 1,\cdots,N. \nonumber
\end{eqnarray}
Then for the scalar products one obtains
\begin{eqnarray}
\label{scalar_products}
{\mathbf H}_1.{\mathbf S}_1 &=& H_1\left[\sin \varphi_1^h \sin \varphi_1 \cos\left(\theta_1^h-\theta_1\right)+\right. \nonumber\\
&& \left. \cos\varphi_1^h \cos\varphi_1 \right]\equiv H_1 \cos\psi_1, \nonumber\\
{\mathbf H}_N.{\mathbf S}_N &=& H_N\left[\sin \varphi_N^h \sin \varphi_N \cos\left(\theta_N^h-\theta_N\right)+\right. \nonumber \\ 
&& \left. \cos\varphi_N^h \cos\varphi_N \right]\equiv H_N\cos\psi_N,
\nonumber \\
{\mathbf S}_i.{\mathbf S}_{i+1} &=& \sin \varphi_i \sin \varphi_{i+1} \cos\left(\theta_i-\theta_{i+1}\right)+ \nonumber \\
&& \cos\varphi_i \cos\varphi_{i+1}\equiv \cos\phi_i,
\end{eqnarray}
where the angle $\phi_i$, $i=1,\cdots,N-1$ is between the spins ${\mathbf S}_i$ and ${\mathbf S}_{i+1}$, and the angles $\psi_1$ and $\psi_N$ are between the vectors ${\mathbf H}_1$ and ${\mathbf S}_1$, and the vectors ${\mathbf H}_N$ and ${\mathbf S}_N$, respectively. 

The free energy $-\beta  F_N$ of this system is 
\begin{equation}
\label{eq:free_energy}
\exp \left(-\beta  F_N\right)=\int_{0}^{2\pi}\prod_{i=1}^{N} \frac{d\theta_i}{4\pi} \int_{0}^{\pi}\prod_{i=1}^{N} d\varphi_i \sin \varphi_i \;\exp \left(-\beta  {\cal H}\right),
\end{equation}
where the normalization is over the solid angle $4\pi$ because
\begin{equation}
\label{solid_angle}
\int_{0}^{2\pi}d\theta\int_{0}^{\pi}d\varphi\;                            \sin \varphi=4\pi.
\end{equation}
In order to perform the integrations we use the expansion 
\begin{equation}
\label{eq:id_harmonics}
e^{z\cos\theta}=\sqrt{\frac{\pi}{2z}}\sum_{n=0}^{\infty}(2n+1)I_{n+1/2}(z) P_n(\cos\theta)
\end{equation}
combined with the addition theorem for the spherical harmonics \cite{AS,NIST2010}
\begin{eqnarray}
\label{eq:id_adding_harmonics}
P_n(\cos \phi_i)&=&\frac{4\pi}{2n+1}\times  \\
&& \sum_{m=-n}^{n} Y_{n,m}^*(\varphi_{i+1},\theta_{i+1})Y_{n,m}(\varphi_i,\theta_i). \nonumber
\end{eqnarray}
Here $I_{n+1/2}(z)$ is the modified Bessel function of the first kind, $P_n(x)$ is the Legendre polynomial of degree $n$ and $Y_{n,m}(\phi,\theta)$ is the spherical harmonic. We remind the orthogonality relation that  holds for the spherical harmonics  
\begin{eqnarray}
\label{eq:Y_ort_prop}
\int_{0}^{\pi}d\varphi \int_0^{2\pi}d\theta\; \sin\varphi\; Y_{l_1,m_1}(\varphi,\theta)Y_{l_2,m_2}^{*}(\varphi,\theta) && \nonumber \\
=\delta_{l_1,l_2}\delta_{m_1,m_2}. &&
\end{eqnarray}
From \eq{eq:free_energy} we obtain 
\begin{eqnarray}
\label{eq:free_energy_calcul}
\exp \left(-\beta  F_N\right) &=& \int_{0}^{2\pi}\prod_{i=1}^{N} \frac{d\theta_i}{4\pi} \int_{0}^{\pi}\prod_{i=1}^{N} d\varphi_i \sin \varphi_i\\
&&\times e^{h_1\cos\psi_1}\left(\prod_{i=1}^{N-1} e^{K\cos\phi_i}\right)e^{h_N\cos\psi_N}, \nonumber
\end{eqnarray}
where $K$, $h_1$ and $h_N$ are defined in accord with \eq{eq:def_parameters_Heis}. 
Now we have to take into account that, according to Eqs. (\ref{eq:id_harmonics}) and (\ref{eq:id_adding_harmonics}),
\begin{widetext}
\begin{eqnarray}
\label{h1}
e^{h_1\cos\psi_1} &=& \sqrt{\frac{\pi}{2h_1}}
 \sum_{n_1=0}^{\infty}(2n_1+1)I_{n_1+1/2}(h_1) P_{n_1}(\cos\psi_1)\\
&=& (4\pi)\sqrt{\frac{\pi}{2h_1}}
\times \sum_{n_1=0}^{\infty} I_{n_1+1/2}(h_1)  \sum_{m_1=-n_1}^{n_1}Y_{n_1,m_1}^{*}(\varphi_1,\theta_1)Y_{n_1,m_1}(\varphi_1^h,\theta_1^h),\nonumber
\end{eqnarray}
\begin{eqnarray}
\label{hN}
e^{h_N\cos\psi_1} &=& \sqrt{\frac{\pi}{2h_N}}
 \sum_{n_N=0}^{\infty}(2n_N+1)I_{n_N+1/2}(h_N) P_{n_N}(\cos\psi_N)\\
&=& (4\pi)\sqrt{\frac{\pi}{2h_N}}
\times \sum_{n_N=0}^{\infty} I_{n_N+1/2}(h_N)  \sum_{m_N=-n_N}^{n_N}Y_{n_N,m_N}^{*}(\varphi_N^h,\theta_N^h) Y_{n_N,m_N}(\varphi_N,\theta_N),\nonumber
\end{eqnarray}
and
\begin{eqnarray}
\label{KN}
e^{K\cos\phi_i}&=&\sqrt{\frac{\pi}{2K}} \sum_{n_{i+1}=0}^{\infty}(2n_{i+1}+1)I_{n_{i+1}+1/2}(K) P_{n_{i+1}}(\cos\phi_{i})\\
&=& (4\pi) \sqrt{\frac{\pi}{2K}}\sum_{n_{i+1}=0}^{\infty}I_{n_{i+1}+1/2}(K)  \sum_{m_{i+1}=-n_{i+1}}^{n_{i+1}}Y_{n_{i+1},m_{i+1}}^{*}(\varphi_{i+1},\theta_{i+1})Y_{n_{i+1},m_{i+1}}(\varphi_{i},\theta_{i}),
\end{eqnarray}
\end{widetext}
with $i=1,\cdots,N-1$. Inserting the above expression into \eq{eq:free_energy_calcul} one can easily perform the integration over $\varphi_i$ and $\theta_i$, $i=1,\cdots,N$ taking into account the orthogonality relations \eq{eq:Y_ort_prop}. One derives that $n_1=n_2=\cdots=n_N=n$, and $m_1=m_2=\cdots=m_N=m$ and, thus, from \eq{eq:free_energy_calcul} we obtain 
\begin{widetext}
\begin{eqnarray}
\label{eq:free_energy_calcul_final}
\lefteqn{\exp \left(-\beta  F_N\right) = (4\pi) \sqrt{\frac{\pi}{2h_1}} \sqrt{\frac{\pi}{2h_N}} \left(\sqrt{\frac{\pi}{2K}}\,\right)^{N-1}}\\
&& \sum_{n=0}^{\infty} I_{n+1/2}(h_1) I_{n+1/2}(h_N) \left[I_{n+1/2}(K)\right]^{N-1}
\sum_{m=-n}^{n} Y_{n,m}(\varphi_1^h,\theta_1^h) Y_{n,m}^{*}(\varphi_N^h,\theta_N^h)\nonumber \\
&=& \left(\frac{\pi}{2K}\right)^{(N-1)/2}\frac{\pi}{2\sqrt{h_1h_N}} \sum_{n=0}^{\infty} (2n+1)P_n \left(\cos\psi_h\right) I_{n+1/2}(h_1) I_{n+1/2}(h_N) \left[I_{n+1/2}(K)\right]^{N-1}, \nonumber
\end{eqnarray}
\end{widetext}
where, in the last line, we have again used the addition theorem for the spherical harmonics  \eq{eq:id_adding_harmonics}. In \eq{eq:free_energy_calcul_final} $\psi_h$ is the angle between the vectors ${\mathbf H}_1$ and ${\mathbf H}_N$ where
\begin{equation}
\label{eq:angle_psi}
\cos\psi_h=\sin \varphi_1^h \sin \varphi_N^h \cos\left(\theta_1^h-\theta_N^h\right)+\cos\varphi_1^h \cos\varphi_N^h.
\end{equation} 

From \eq{eq:free_energy_calcul_final} and \eq{eq:free_energy_calcul_final_result} for the total pressure exerted by the end points on the system one derives 
\begin{widetext}
\begin{equation}
\label{eq:def_Ftot_Heis}
\beta  F_{\text{tot}}\equiv -\frac{\partial}{\partial N}[\beta  F_N]=\ln\left[\frac{\sinh K}{K}\right]+\frac{ \sum_{n=1}^{\infty} (2n+1)P_n \left(\cos\psi_h\right) \ln\left[\frac{I_{n+1/2}(K)}{I_{1/2}(K)}\right] \frac{I_{n+1/2}(h_1)}{I_{1/2}(h_1)} \frac{I_{n+1/2}(h_N)}{I_{1/2}(h_N)} \left[\frac{I_{n+1/2}(K)}{I_{1/2}(K)} \right]^{N-1}}{1+ \sum_{n=1}^{\infty} (2n+1)P_n \left(\cos\psi_h\right) \frac{I_{n+1/2}(h_1)}{I_{1/2}(h_1)} \frac{I_{n+1/2}(h_N)}{I_{1/2}(h_N)} \left[\frac{I_{n+1/2}(K)}{I_{1/2}(K)}\right]^{N-1}}.
\end{equation}
\end{widetext}
From here one derives the exact result for the Casimir force reported in \eq{eq:def_FCas_Heis} in the main text. From it one can extract the corresponding scaling behavior reported in  \eq{eq:def_FCas_Heis_ass} which is convenient for evaluation of the behavior of the force for moderate and large values of the scaling variable $x$. Here we present the corresponding derivation of the representation convenient for extracting the behavior of the force for small values of the scaling variable. Let us start by considering the sum 
\begin{widetext}
\begin{eqnarray}
\label{smallx_Heis}
S(\psi_h,h_1,h_N,K) &\equiv& \sum_{n=1}^{\infty} (2n+1)P_n \left(\cos\psi_h\right) \frac{I_{n+1/2}(h_1)}{I_{1/2}(h_1)} \frac{I_{n+1/2}(h_N)}{I_{1/2}(h_N)} \left[\frac{I_{n+1/2}(K)}{I_{1/2}(K)} \right]^{N-1} \nonumber  \\
 & \simeq & \sum_{n=1}^{\infty} (2n+1)P_n \left(\cos\psi_h\right) \frac{e^{h_1-(n+1/2)^2/(2h_1)}}{\sqrt{2\pi h_1}I_{1/2}(h_1)}\frac{e^{h_N-(n+1/2)^2/(2h_N)}}{\sqrt{2\pi h_N}I_{1/2}(h_N)}
 \left[\frac{e^{K-(n+1/2)^2/(2K)}}{\sqrt{2\pi K}I_{1/2}(K)}\right]^{N-1}
 \nonumber \\
 &\simeq& 
  \sum_{n=1}^{\infty} (2n+1)P_n \left(\cos\psi_h\right) 
 \frac{\exp{\left[-\frac{1}{2}\left(n+1/2\right)^2\left(\frac{1}{h_1}+\frac{1}{h_N}+\frac{N-1}{K}\right)\right]}}{\exp{\left[-\frac{1}{2}\left(1/2\right)^2\left(\frac{1}{h_1}+\frac{1}{h_N}+\frac{N-1}{K}\right)\right]}}
 \nonumber \\
   &\simeq &
     \sum_{n=1}^{\infty} (2n+1)P_n \left(\cos\psi_h\right) 
      \frac{I_{n+1/2}\left(\frac{1}{h_{\rm eff}^{-1}+x}\right)}{I_{1/2}\left(\frac{1}{h_{\rm eff}^{-1}+x}\right)}=\sum_{n=0}^{\infty} (2n+1)P_n \left(\cos\psi_h\right) 
            \frac{I_{n+1/2}\left(\frac{1}{h_{\rm eff}^{-1}+x}\right)}{I_{1/2}\left(\frac{1}{h_{\rm eff}^{-1}+x}\right)}-1
      \nonumber \\
      &=& \sqrt{\frac{2}{\pi \left(h_{\rm eff}^{-1}+x\right)} }\frac{\exp{\left[\frac{\cos\psi_h}{h_{\rm eff}^{-1}+x}\right]}}{I_{1/2}\left(\frac{1}{h_{\rm eff}^{-1}+x}\right)}-1=\frac{1}{ \left(h_{\rm eff}^{-1}+x\right)} \frac{\exp{\left[\frac{\cos\psi_h}{h_{\rm eff}^{-1}+x}\right]}}{\sinh\left(\frac{1}{h_{\rm eff}^{-1}+x}\right)}-1. 
\end{eqnarray}
\end{widetext}

\section{Calculation of the free energy for the 3d Gaussian model}
\label{A:GM}

In the current appendix we will outline some technical steps needed to obtain the free energy of the Gaussian model under the considered boundary conditions. 

Performing the Fourier transform 
\begin{multline}
\label{eq:Fourier}
S_{x,y,z}=\frac{1}{\sqrt{{M}}} \sum _{m=1}^M \left[\cos
   \left(\frac{2 \pi}{M} m x\right)+\sin
      \left(\frac{2 \pi}{M} m x\right)\right] \\
      \times \frac{1}{\sqrt{N}} \sum _{n=1}^N \left[\cos
         \left(\frac{2 \pi}{N} n y\right)+\sin
                  \left(\frac{2 \pi}{N} n y\right)\right] \;\;\;\;\;\\
\times \sqrt{\frac{2}{L+1}} \sum _{l=1}^L  \sin
   \left(\frac{\pi}{L+1} l z\right)
   \tilde{S}_{m,n,l} 
\end{multline}
in Eq. \eqref{eq:def_Ham_GM_final}, one can easily diagonalize the Hamiltonian.  Then, performing the integrations over $\tilde{S}_{m,n,l}$, $m=1,\cdots,M$, $n=1,\cdots,N$ and $l=1,\cdots,L$ one immediately obtains Eqs. (\ref{eq:free_energy_GM}) and (\ref{eq:free_energy_h}) for the filed-independent and field-dependent parts of the free energy reported in the main text. In what follows we explain how to perform the summations in these terms. We start with the term that depends on the applied surface fields.

\subsection{Evaluation of the field dependent term}
Taking $L$, for definiteness, to be odd number, we start by rewriting \eq{eq:free_energy_GM} in the form
\begin{equation}
\label{eq:odd}
\Delta F_h=\Delta F_h^{\rm odd}+\Delta F_h^{\rm even},
\end{equation}
where

i) if $p\ne  M$ or $q\ne N$:
\begin{eqnarray}
\label{eq:free_energy_even}
-\beta \Delta F_h^{\rm even} &=&  \frac{MN}{8(L+1)K^{\perp}}\; S^{\rm even}(\Lambda,L) \times  \\
            && \left[h_1^2+h_L^2-2 h_L h_1
                            \cos(\mathbf{k.\Delta})\right], \nonumber
\end{eqnarray}
and
\begin{eqnarray}
\label{eq:free_energy_odd}
-\beta \Delta F_h^{\rm odd} &=&  \frac{MN}{8(L+1)K^{\perp}}\; S^{\rm odd}(\Lambda,L) \times \\
            &&\left[h_1^2+h_L^2+2 h_L h_1
                            \cos(\mathbf{k.\Delta})\right], \nonumber
\end{eqnarray}

ii) if $p=M$ and $q=N$ 
\begin{eqnarray}
\label{eq:free_energy_even}
-\beta \Delta F_h^{\rm even} &=&  \frac{MN}{8(L+1)K^{\perp}}\; S^{\rm even}(\Lambda,L) \times  \\
            && \left[h_1-h_L
                                       \cos 2 \pi  ( \Delta
                                       _x+ \Delta
                                       _y)\right]^2, \nonumber
\end{eqnarray}
and
\begin{eqnarray}
\label{eq:free_energy_odd}
-\beta \Delta F_h^{\rm odd} &=&  \frac{MN}{8(L+1)K^{\perp}}\; S^{\rm odd}(\Lambda,L) \times \\
            && \left[h_1+h_L
                           \cos 2 \pi  ( \Delta
                           _x+ \Delta
                           _y)\right]^2. \nonumber
\end{eqnarray}

In the above expressions 
\begin{equation}
\label{eq:S_even_def}
S^{\rm even}(\Lambda,L)=\sum _{l=1}^{(L-1)/2} \frac{\sin^2\left(\frac{2\pi l}{L+1}\right)}{\Lambda- \cos\left(\frac{2\pi l}{L+1}\right)},
\end{equation}
\begin{equation}
\label{eq:S_odd_def}
S^{\rm odd}(\Lambda,L)=\sum _{l=1}^{(L-1)/2} \frac{\sin^2\left(\frac{\pi(2l+1) }{L+1}\right)}{\Lambda- \cos\left(\frac{\pi (2l+1)}{L+1}\right)},
\end{equation}
and
\begin{eqnarray}
\label{eq:Lambda_def}
\Lambda&=& 1+\left(\frac{\beta_c}{\beta}-1\right)\left[ 2\frac{J^{\|}}{J^\perp}+1\right]\nonumber \\
         &&+\dfrac{J^{\|}}{J^{\perp}}\left[2-\cos \left(\frac{2 \pi 
               p}{M}\right)-\cos \left(\frac{2 \pi 
                  q}{N}\right)\right].
\end{eqnarray}
It is easy to show that 
\begin{equation}
\label{eq:S_even}
S^{\rm even}(\Lambda,L)= \frac{1}{2} (L-1) \Lambda+(1-\Lambda^2){\hat S}^{\rm even}(\Lambda,L)
\end{equation}
where
\begin{equation}
\label{eq:hat_S_even}
{\hat S}^{\rm even}(\Lambda,L)=\sum _{l=1}^{(L-1)/2} \frac{1}{\Lambda- \cos\left(\frac{2\pi l}{L+1}\right)}
\end{equation}
and that
\begin{equation}
\label{eq:S_odd}
S^{\rm odd}(\Lambda,L)= \frac{1}{2} (L+1) \Lambda+(1-\Lambda^2){\hat S}^{\rm odd}(\Lambda,L)
\end{equation}
where
\begin{equation}
\label{eq:hat_S_odd}
{\hat S}^{\rm odd}(\Lambda,L)=\sum _{l=1}^{(L-1)/2} \frac{1}{\Lambda- \cos\left(\frac{\pi (2l+1)}{L+1}\right)}.
\end{equation}
The summations in \eq{eq:hat_S_even} and \eq{eq:hat_S_odd} can be performed using \cite{GR} the identities 
\begin{equation}
\label{eq:identity1}
\cosh n x - \cos n y =2^{n-1}\prod_{k=0}^{n-1} \left[\cosh x -\cos \left(y+\dfrac{2\pi k}{n}\right)\right]
\end{equation}
and 
\begin{equation}
\label{eq:identity2}
\cos n x - \cos n y =2^{n-1}\prod_{k=0}^{n-1} \left[\cos x -\cos \left(y+\dfrac{2\pi k}{n}\right)\right].
\end{equation}
With the help of the variable $\lambda$, introduced in \eq{eq:x_def_1} and \eq {eq:x_def_2},
 for the sums ${\hat S}^{\rm even}(\Lambda,L)$ and ${\hat S}^{\rm odd}(\Lambda,L)$ we obtain 
\begin{eqnarray}
\label{eq:S_even_result_1}
&&{\hat S}^{\rm even}(\Lambda,L)=\frac{\Lambda }{1-\Lambda ^2}\\
&&+\frac{1}{2} (1+L) \coth
   \left[\frac{1}{2} (1+L) \lambda\right] \cosh(\lambda), \qquad \Lambda\ge 1.\nonumber
\end{eqnarray}
and
\begin{eqnarray}
\label{eq:S_even_result_2}
&&{\hat S}^{\rm even}(\Lambda,L)=\frac{\Lambda }{1-\Lambda ^2}\\
&&-\frac{1}{2} (1+L) \cot
   \left[\frac{1}{2} (1+L) \lambda\right] \csc(\lambda), \qquad \Lambda\le 1.\nonumber
\end{eqnarray}
for ${\hat S}^{\rm even}(\Lambda,L)$, while for the sum ${\hat S}^{\rm odd}(\Lambda,L)$ one has
\begin{eqnarray}
\label{eq:S_odd_result_1}
{\hat S}^{\rm odd}(\Lambda,L)= \frac{1}{2} (1+L) \dfrac{\tanh
   \left[\frac{1}{2} (1+L) \lambda\right]}{\sinh \lambda}, \Lambda\ge 1,\;\;\;\;\;\;
\end{eqnarray}
and 
\begin{eqnarray}
\label{eq:S_odd_result_2}
{\hat S}^{\rm odd}(\Lambda,L)= \frac{1}{2} (1+L) \dfrac{\tan
   \left[\frac{1}{2} (1+L) \lambda\right]}{\sin \lambda}, \Lambda\le 1.\;\;\;
\end{eqnarray}
 Obviously, the two  pairs  \eq{eq:S_even_result_1} and \eq{eq:S_even_result_2}, and \eq{eq:S_odd_result_1} and \eq{eq:S_odd_result_2} represent a continuation from real to purely complex values of $\lambda$. Because of that, in the remainder we will report only one of the corresponding representations concerning the sums. 

From \eq{eq:S_even} and \eq{eq:S_even_result_1} one obtains
\begin{equation}
\label{eq:S_even_final}
S^{\rm even}(\Lambda,L)=\dfrac{L+1}{2} \left\{ \Lambda-\coth\left[\dfrac{L+1}{2} \lambda\right] \sinh[\lambda]\right\}
\end{equation}
whereas from \eq{eq:S_odd} and \eq{eq:S_odd_result_1} one derives 
\begin{equation}
\label{eq:S_odd_final}
S^{\rm odd}(\Lambda,L)=\dfrac{L+1}{2} \left\{ \Lambda-\tanh\left[\dfrac{L+1}{2} \lambda\right] \sinh[\lambda]\right\}.
\end{equation}
Using the above expressions and taking into account Eqs. \eqref{eq:odd} - \eqref{eq:free_energy_odd} for $\Delta f_h$, see \eq{eq:f_h},
one obtains

i) if $p\ne  M$ or $q\ne N$:
\begin{eqnarray}
\label{eq:delta_fh_final_pq}
-\beta \Delta f_h &=& \frac{1}{16 K^{\perp}} \left\lbrace
\left[h_1^2+h_L^2-2 h_L h_1
                            \cos(\mathbf{k.\Delta})\right] \right. \nonumber  \\
&& \left. \times \left[ \Lambda-\coth\left[\dfrac{L+1}{2} \lambda\right] \sinh[\lambda]\right]\right. 
\nonumber\\
&&\left.+
\left[h_1^2+h_L^2+2 h_L h_1
                            \cos(\mathbf{k.\Delta})\right]\right. \nonumber\\
&&\left. \times\left[ \Lambda-\tanh\left[\dfrac{L+1}{2} \lambda\right] \sinh[\lambda]\right]
 \right\rbrace,
\end{eqnarray}

ii) if $p=M$ and $q=N$ 

\begin{eqnarray}
\label{eq:delta_fh_final}
-\beta \Delta f_h &=& \frac{1}{16 K^{\perp}} \left\lbrace
\left[h_1-h_L
                                       \cos 2 \pi  ( \Delta
                                       _x+ \Delta
                                       _y)\right]^2 \right. \nonumber  \\
&& \left. \times \left[ \Lambda-\coth\left[\dfrac{L+1}{2} \lambda\right] \sinh[\lambda]\right]\right. 
\nonumber\\
&&\left.+
\left[h_1+h_L
                                       \cos 2 \pi  ( \Delta
                                       _x+ \Delta
                                       _y)\right]^2 \right. \nonumber\\
&&\left. \times\left[ \Lambda-\tanh\left[\dfrac{L+1}{2} \lambda\right] \sinh[\lambda]\right]
 \right\rbrace,
\end{eqnarray}

Note that in deriving the above expression no approximations have been made - it is an {\it exact} result. 

If $L \lambda\gg 1$ from the above one immediately obtains

i) if $p\ne  M$ or $q\ne N$:
\begin{eqnarray}
\label{eq:large_x_pq}
-\beta \Delta f_h &\simeq& \dfrac{1}{8K^{\perp}}  \left\lbrace \Lambda-\sinh[\lambda]\right\rbrace \left\lbrace h_1^2+h_L^2\right\rbrace \nonumber\\
&&+\dfrac{1}{2 K^{\perp}}  \sinh[\lambda] e^{-(L+1)\lambda} h_1 h_L \cos\left(\mathbf{k.\Delta}\right),\ \ \ \ \ \ 
\end{eqnarray}
and 

ii) if $p=M$ and $q=N$ 
\begin{eqnarray}
\label{eq:large_x}
\lefteqn{-\beta \Delta f_h \simeq} \\
&& \dfrac{1}{8K^{\perp}}  \left\lbrace \Lambda-\sinh[\lambda]\right\rbrace \left\lbrace h_1^2+\cos 2 \pi  ( \Delta
                                       _x+ \Delta
                                       _y) h_L^2\right\rbrace \nonumber\\
&&+\dfrac{1}{2 K^{\perp}}  \sinh[\lambda] e^{-(L+1)\lambda} h_1 h_L \cos 2 \pi  ( \Delta
                                       _x+ \Delta
                                       _y),\nonumber 
\end{eqnarray}
wherefrom one derives the surface part $\Delta f_h^{(s)}$ of the field-dependent term in the free energy 

i) if $p\ne  M$ or $q\ne N$:
\begin{equation}
\label{eq:bulk_h}
-\beta \Delta f_h^{(s)} = \dfrac{1}{8K^{\perp}}  \left\lbrace \Lambda-\sinh[\lambda]\right\rbrace \left\lbrace h_1^2+h_L^2\right\rbrace. 
\end{equation}
and 

ii) if $p=M$ and $q=N$ 
\begin{eqnarray}
\label{eq:bulk_h}
-\beta \Delta f_h^{(s)} & = & \dfrac{1}{8K^{\perp}}  \left\lbrace \Lambda-\sinh[\lambda]\right\rbrace \nonumber \\
&& \times \left\lbrace h_1^2+\cos 2 \pi (\Delta_x+ \Delta_y) h_L^2\right\rbrace. 
\end{eqnarray}

From Eqs. \eqref{eq:delta_fh_final_pq} and \eqref{eq:delta_fh_final} one can determine both the transverse and the longitudinal field contribution to the components of the Casimir force. The corresponding results are reported in the main text.

\subsection{Evaluation of the field independent term}

We are interested in the $L$-dependent behavior of the field-independent part of the statistical sum of the system, see \eq{eq:f_0}, 
where $\Delta F_0$ is given by \eq{eq:free_energy_GM}. It is easy to see that 
\begin{eqnarray}
\label{eq:f0_int}
\lefteqn{-\beta \Delta f_0-\frac{1}{2} L \ln \frac{\pi}{K^{\perp}}=-\frac{1}{2} \frac{1}{(2\pi)^2}\int_0^{2\pi}d\theta_1\int_0^{2\pi}d\theta_2  } \;\;\;\;\;\;\; \\
&=&-\frac{1}{2} \frac{1}{(2\pi)^2}\int_0^{2\pi}d\theta_1\int_0^{2\pi}d\theta_2\left.  S_0\left(\frac{\beta _c}{\beta
   },\frac{J^\|}{J^\perp},L \right| \theta
   _1,\theta _2\right) \nonumber
\end{eqnarray}
where 
\begin{eqnarray}
\label{S0def}
\lefteqn{\left. S_0\left(\frac{\beta _c}{\beta},\frac{J^\|}{J^\perp},L\right|\theta
   _1,\theta _2\right)\equiv}\\
   &&\sum_{k=1}^L \ln \left[\frac{s}{K^{\perp}} -\frac{K^\|}{K^\perp}\left(\cos \theta_1+\cos \theta_2\right)-\cos \frac{\pi k}{L+1}\right] \nonumber\\
   &=& \sum_{k=1}^L \ln \left[\left(\frac{\beta_c}{\beta}-1\right)\left(1+ 2\frac{J^{\|}}{J^\perp}\right) +\left(1-\cos \frac{\pi k}{L+1}\right)\right. \nonumber\\
   && \left. +\frac{J^\|}{J^\perp}\left(2-\cos \theta_1-\cos \theta_2\right)\right] \nonumber
\end{eqnarray}
and we have used \eq{betac}. 

The expression in \eq{eq:f0_int} can be evaluated in several ways. Let us briefly sketch one of them. By doing so we will also obtain an expression for the free energy that has not been derived before and which are valid not only for large, but for any positive value of $L$.  

Using the identity in \eq{eq:identity1} one can show that 
\begin{equation}
\label{eq:S0}
\left. S_0\left(\frac{\beta _c}{\beta},\frac{J^\|}{J^\perp},L\right|\theta
   _1,\theta _2\right)=-L \ln 2+\ln
   \left[\frac{\sinh (1+L)
   \delta}{\sinh \delta}\right],
\end{equation}
where $\delta$ is defined in \eq{eq:def_delta}. 
For the  contribution of the field-independent term to the transverse Casimir force $\beta\Delta F^{(0,\perp)}_{\rm Cas}$, see \eq{eq:_def_no_field},
from \eq{eq:f0_int} and \eq{eq:S0} one derives \eq{eq:Cas_no_field} given in the main text. 
In order to derive the scaling form of $\Delta F^{(0,\perp)}_{\rm Cas}$ we have to consider the regime $L\gg 1$. 
Obviously, then Casimir force will be exponentially small if $\delta$ is finite. In order to avoid that, one needs $\delta\to 0$ so that $(L+1)\delta=O(1)$. When $\delta$ goes to zero, however, both $(\beta_c/\beta-1)(1+2 J^\|/J^\perp)\to 0$ and $\theta_1, \theta_2\to 0$. Then, from \eq{eq:def_delta} one obtains 
\begin{equation}
\label{eq:delta}
\delta^2=2\left(\frac{\beta_c}{\beta}-1\right)\left(1+ 2\frac{J^{\|}}{J^\perp}\right) +\frac{J^\|}{J^\perp}\left( \theta_1^2+\theta_2^2\right). 
\end{equation}
Passing to polar coordinates, from \eq{eq:Cas_no_field} one obtains, up to exponentially small in $L$ corrections  
\begin{equation}
\label{eq:Cas_no_field_ev}
\beta\Delta F^{(0,\perp)}_{\rm Cas}=-\frac{1}{2}
   \int _{\delta_{\rm min} }^{\infty }x^2
   \left[\coth ((1+L) x)-1\right] \frac{dx}{2\pi} 
\end{equation}
where
\begin{equation}
\label{eq:delta_min}
\delta_{\rm min}=\sqrt{2\left(\frac{\beta_c}{\beta}-1\right)\left(1+ 2\frac{J^{\|}}{J^\perp}\right)}.
\end{equation}
Noting that
\begin{equation}
\label{delta_xt}
x_t=L\delta_{\rm min}, 
\end{equation}
using that
\begin{equation}
\label{eq:coth}
\coth x= 1+2 \sum _{k=1}^{\infty } e^{-2 k x}
\end{equation}
and performing the integration in \eq{eq:Cas_no_field_ev}, one derives Eqs. (\ref{eq:F_Cas_no_field}) and (\ref{eq:X_Cas_no_field_sf}) given in the main text.
From \eq{eq:Cas_no_field_ev} and taking into account the definition \eqref{delta_xt} one immediately concludes that $X^{(0,\perp)}_{\rm Cas}(x_t)$ is a {\it monotonically increasing} function of $x_t$.

%


\end{document}